\documentclass[epjc3]{svjour3}

\RequirePackage[T1]{fontenc}

\smartqed  

\RequirePackage{mathptmx}      
\RequirePackage{flushend}
\RequirePackage[numbers,sort&compress]{natbib}
\RequirePackage[colorlinks,citecolor=blue,urlcolor=blue,linkcolor=blue]{hyperref}

\journalname{Eur. Phys. J. C}

\usepackage{enumerate}

\usepackage{amsmath,amstext,amsfonts,amscd,latexsym,amssymb,amstext,mathrsfs,dsfont,xfrac}

\usepackage{graphicx,color,pst-plot}
\usepackage{pgfplots,tikz}
\usetikzlibrary{snakes}

\allowdisplaybreaks[1]

\def\Beta{{\rm \scriptstyle B}}
\def\R{{\rm Re\,}}

\def\e{{\rm e\,}}
\def\d{{\rm d }}
\newcommand{\hyper}[5]{\;_{#1}{\rm F}_{#2} \left(\left.\begin{matrix} {#3} \\ {#4} \end{matrix}\right| #5\right) }

\begin{document}

\title{Assuming Regge trajectories in holographic QCD: from OPE to Chiral Perturbation Theory}

\author{Luigi Cappiello \thanksref{e1,addr1,addr2} \and Giancarlo D'Ambrosio \thanksref{e2,addr2,addr3} \and David Greynat \thanksref{e3,addr1,addr2}.}

\thankstext[$\star$]{t1}{Preprint number: CERN-PH-TH-2015-107}
\thankstext{e1}{e-mail: luigi.cappiello@na.infn.it}
\thankstext{e2}{e-mail: gdambros@na.infn.it}
\thankstext{e3}{e-mail: david.greynat@gmail.com}

\institute{Dipartimento di Scienze Fisiche, Universit\'a di Napoli "Federico II", Via Cintia, 80126 Napoli, Italia \label{addr1}\\
\and INFN-Sezione di Napoli, Via Cintia, 80126 Napoli, Italia\label{addr2}\\
\and CERN Theory Division, CH-1211 Geneva 23, Switzerland \label{addr3} }

\date{Received: date / Accepted: date}

\maketitle

\begin{abstract}
The soft wall model in holographic QCD has Regge trajectories but wrong operator product expansion (OPE) for the two-point vectorial QCD Green function. We modify the dilaton potential to comply with the OPE. We study also the axial two-point function using the same modified dilaton field and an additional scalar field to address chiral symmetry breaking. OPE is recovered adding a boundary term and low energy chiral parameters, $F_\pi$ and $L_{10}$, are well described analytically by the model in terms of Regge spacing and QCD condensates. The model nicely supports and extends previous theoretical analyses advocating Digamma function to study QCD two-point functions in different momentum regions.
\end{abstract}

\section{Introduction}

QCD Green functions describe three ranges of energies \textit{(i)} deep euclidean, where perturbative QCD and OPE expansion  methods can be applied, \textit{(ii)} an intermediate Minkowski region, where resonances are described by Regge trajectories  and \textit{(iii)} the strong interacting, low-energy region, described by chiral perturbation theory ($\chi$PT)  \cite{Peris:1998nj}.

Regge trajectories, \textit{i.e.} a linear growth of the square of the resonance masses like $M_V ^2 (n)= \sigma n$ GeV$^2$ for the vectors, were conjectured long ago,  before QCD, and very well verified phenomenologically \cite{collins}. This relation, thought to be valid in QCD for large number of colours, $N_c$, has been proven by 't Hooft only in 2 dimensions \cite{tHooft2d}. Also other arguments have been proposed for the validity  of Regge theory aside from the experimental evidence: lattice,  flux tube, confinement  \cite{Golterman:2001nk,Kaidalov:2001db,Shifman:2007xn}: we insist on these theoretical motivations of Regge trajectories  through this paper since  we believe that QCD dynamics favours this physical picture.
 
Low energy QCD properties, like the chiral symmetry breaking ($\chi$SB) parameters, $F_\pi$  and the Gasser-Leutwyler coefficients $L_i$'s of the ${\cal O}(p^4)$ chiral Lagrangian have been studied in the framework of holographic models of QCD, based on the  AdS/CFT correspondence \cite{Maldacena:1997re, Gubser:1998bc, Witten:1998qj}. Several five dimensional set-ups have been proposed. The basic feature of Hard-Wall (HW) models  is to simulate confinement by cutting drastically the extradimension of the AdS$_5$ in the IR, producing an infinite spectrum of Kaluza-Klein (KK) states  to be identified with vector resonances of increasing masses. $\chi$SB in the axial sector is triggered either  by  a scalar field in the bulk \cite{Erlich:2005qh, Da Rold:2005zs} or by  appropriate boundary conditions \cite{Hirn:2005nr}. These models were actually anticipated by a deconstruction model   \cite{Son:2003et}.  Also  the  Sakai-Sugimoto model \cite{Sakai:2004cn}, a string set-
 up, shares this feature  of an infinite spectrum of KK states. A recent review of all these models and their relation with the light-front holographic QCD approach is Ref \cite{Brodsky:2014yha}.

All these models  have a good description of the deep euclidean region; for instance the correct two point function recover the partonic log, and a pretty well description of low energy QCD, obtaining the chiral parameters $F_\pi$ and the Gasser-Leutwyler coefficients  $L_i$'s which are close to the physical values. In fact, it was observed that these HW models \cite{Erlich:2005qh, Da Rold:2005zs,Hirn:2005nr} have the same two point vectorial Green function as the one obtained by Migdal long ago who proposed an \textit{ad hoc} prescription to perform the analytic continuation of the perturbative deep Euclidean QCD result to the Minkowski region \cite{Shifman:2005zn,Migdal:1977ut}.

In order to reproduce resonances masses with a Regge spacing,  one can consider a 5D model with AdS metric and an additional field, the dilaton \cite{Karch:2006pv}. However, in this model it can be shown that the partonic log of the two point vectorial Green function
\begin{equation}
\Pi_V(-q^2) = - \sum_{n=0}^\infty \frac{F_V(n)^2}{q^2-M_V(n)^2} \underset{Q^2 \rightarrow \infty}{\thicksim}  \frac{N_c}{24\pi^2}\ln \left(\frac{\Lambda_V^2}{Q^2}\right)
\end{equation}
with the Euclidean momentum $Q^2 \doteq -q^2$, receives $\sfrac{1}{Q^2}$-corrections differently from what holds in perturbative QCD Operator Product Expansion (OPE)\footnote{Due to its definition, $\Pi_V$ used in the holographic models \cite{Erlich:2005qh} differs by a factor $\sfrac{1}{2}$ from the $\Pi_V$ used in \cite{Peris:1998nj,Golterman:2001nk}.} \cite{Andreev:2006vy,Cata:2006ak,Zuo:2008re,Csaki:2006ji}. Nevertheless, we find extremely interesting to have a model where the Regge region would be analytically related to the deep Euclidean region even though with the wrong OPE. This will be exactly our starting point: OPE tells us the correct Green functions in the deep Euclidean in terms of gluon and quarks condensates. Is it possible to modify the dilaton profile, $\phi(z)\mapsto \phi(z) + \delta \phi(z) $, such to comply with QCD requirements in the intermediate (Regge) region and UV (OPE) region? Once this question has been answered in the affirmative, we are naturally led to add
 ress another important problem, that does not have a satisfying answer in this "linear confinement" holographic approach \cite{Karch:2006pv}: how to implement the effects of chiral symmetry breaking into our model. If we look at the expansion of both vector-vector and axial-axial correlator $ \Pi_{V,A}(Q^2)$ for large Euclidean momentum, QCD predicts
\begin{equation}
\Pi_{V,A}(Q^2) \underset{Q^2 \rightarrow \infty}{\thicksim}  \frac{N_c}{24\pi^2}\ln \left(\frac{\Lambda_V^2}{Q^2}\right)   +  \frac{\alpha_s  \left\langle G^2\right\rangle }{24\pi}  \frac{1}{Q^4}   -\frac{14 \pi}{9}  c_{V,A} \;\alpha_s\; \left\langle \bar{\psi}\psi\right\rangle^2   \frac{1}{Q^6} 
\end{equation}
with $c_{V}=1$ and $c_{A}=-\sfrac{11}{7}$. We should then be able to take into account the difference appearing at order $\sfrac{1}{Q^6}$ in the OPE expansion of the vector and axial correlators. On the other hand, in the low and intermediate region in Minkowskian momentum, chiral symmetry is broken at the level of the mass spectra of resonances, with the appearance of the pion as the corresponding pseudo-Goldston boson of $\chi$SB. If we were able to deal with these aspects in a holographic model, we would have, in principle, closed the circle, since we would have a description of all three energy ranges of QCD. In fact, implementing the correct OPE and mass spectrum for the axial sector, we will propose our post-diction of $F_\pi$ and $L_{10}$ (one of the the chiral $\mathcal{O}(p^4)$ coefficients) in terms of our input parameters, \textit{i.e.} the Regge spacing and the QCD condensates. Our model, beside being a novel proposal for holographic QCD, links naturally to previous 4D QCD work where a phenomenological matching between low energy and OPE was realised through Regge theory (or Veneziano model) \cite{Peris:1998nj,Veneziano:1968yb,Golterman:2001pj,Golterman:2001nk,Dominguez:2001zu,Afonin:2004yb,Cata:2005zj}.

We have organized this paper in the following way. In Section \ref{sec:1} we summarize the relevant properties of the existing Soft Wall model (SW) and focus on the vectorial two-point function. The issue of $\chi$SB is dealt with in Section \ref{sec:3}. In section \ref{sec:4}, the analytic continuation at low energy is described and the predictions for $F_\pi$ and $L_{10}$ are given as well as some consequences on duality violations. Conclusions are given in section \ref{sec:5}. 

\section{Assumptions and aims for the model}
\label{sec:1}

\subsection{Preliminaries}
\label{sec:Preliminaries}

The Soft-Wall model is a five dimensional model where the additional coordinate, $z$, has the range $0<z<\infty$ and  background fields consist  in a five-dimensional AdS metric and a dilaton field $\Phi(z)$. The AdS metric is written as 
\begin{equation}
\label{eq:ds2}
g_{MN} \; \d x^M \d  x^N = \frac{1}{z^2} \; \left(\eta_{\mu \nu}\d x^{\mu} \d x^{\nu} - \d z^2\right)\;,
\end{equation}
where $\eta_{\mu\nu} = {\rm Diag}\, (1,-1,-1,-1)$, the Greek indices $\mu, \nu = (0,1,2,3)$ referring to the usual 4-dimensions, and the capital Latin ones $M, N = (0,1,2,3,z)$ to the 5 dimensions. It was shown, in \cite{Karch:2006pv}, that with the choice of a quadratic profile of the dilaton field
\begin{equation}
\Phi(z) \doteq \kappa^2 z^2,
\end{equation}
the spectrum of vector resonances followed a Regge trajectory. Vector meson resonances are obtained as the modes associated to a five-dimensional gauge field  $\mathbb{V}_M$ in the external metric and dilaton background, with 5D action 
\begin{equation}
\label{eq:lagragian5DVect}
S_{5} = -\frac{1}{4g_5^2} \int \d^4x \int_0^{\infty} \d z \sqrt{g} \; \e^{-\Phi(z)}\; g^{MN}g^{RS} \; \mathrm{Tr} \left[ \mathbb{F}_{MR} \, \mathbb{F}_{NS} \right]\;,
\end{equation}
with $g = | \det g_{MN} |$,  the field strength $\mathbb{F}_{MN}=\partial_M \mathbb{V}_M-\partial_N \mathbb{V}_M - i \left[\mathbb{V}_M,\mathbb{V}_N \right]$ and $g_5^2=\sfrac{12\pi^2}{N_c}$ is the 5D coupling constant where $N_c$ is the number of colors of QCD. The trace is understood as the sum  over the   $\mathrm{SU}(N_f)$ flavour  indices of $\mathbb{V}_N= t^a \mathbb{V}^a_N$, if $N_f=2$, $t^a=\sfrac{\sigma^a}{2}$ with $\sigma^a$ being Pauli matrices.

We shall work in the axial gauge  $\mathbb{V}_z =0$. The AdS/CFT correspondence prescribes that the boundary value of the 5D gauge field $\mathbb{V}_\mu$ has to be identified with the classical source $\mathsf{v}_\mu$ coupled to the the 4-dimensional vectorial current $J_V^a\ _\mu= :\!\bar{q} \,\gamma_\mu\, t^a q\!:$,
\begin{equation}
\label{eq:vectorialsource}
\lim_{z \rightarrow 0} \mathbb{V}^{a}_{\mu,z}(x,z) = \mathsf{v}^a_\mu(x)  \;.
\end{equation} 

The corresponding equation of motion for the gauge field, derived from the Lagrangian (\ref{eq:lagragian5DVect}), is more easily written in terms of the 4-dimensional Fourier transform $f_V(-q^2,z)$ of the field $\mathbb{V}_M$,
\begin{equation}
f_V(-q^2,z) \; \hat{\mathsf{v}}_\mu(q) \doteq \int  \d^4x  \;  \e^{ - i q \cdot x}  \; \mathbb{V}_{\mu,0}(x,z)\;,
\end{equation}
where $\hat{\mathsf{v}}_\mu$ is the Fourier transform of the source $\mathsf{v}_\mu$. Using gauge invariance, one may assume the $\hat{\mathsf{v}}_\mu$ to be transverse and one obtains 
\begin{equation}
\label{eq:eqmotionvector}
\partial_z^2 \, f_V + \partial_z \left[ \ln w_0(z)\right]  \partial_z \, f_V  + q^2\, f_V = 0 \;,
\end{equation}  
with, 
\begin{equation}
\label{eq:w0}
w_0(z) \doteq \frac{\e^{-\Phi(z)}}{z}  = \frac{\e^{- \kappa^2 z^2}}{z} \;.
\end{equation}

Notice that, in order to satisfy the the boundary condition (\ref{eq:vectorialsource}), we have to impose 
\begin{equation}
\label{eq:fv=1}
f_V(-q^2, 0)=1\;.
\end{equation}

It is well known that the presence of (an infinite set of) solutions of the 5D equation of motion (\ref{eq:eqmotionvector}), satisfying a normalization condition in the extra-dimension $z$, corresponds to (an infinite set of) resonances in 4D. Indeed, the corresponding eigenvalue equation for wave functions $ \phi_n(z) $ of the normalizable modes can be obtained from (\ref{eq:eqmotionvector}) by requiring $ q^2 = M_V(n)^2$, which gives $\phi_n(z) = f_V(-M_V(n)^2,z)$ and the equation
\begin{equation}
\label{eq:eqmotionphi}
\partial_z^2 \, \phi_n - \left( \frac{1}{z} + 2\kappa^2 z \right)  \partial_z \, \phi_n  + M_V(n)^2\, \phi_n = 0 \;,
\end{equation}  
with the vanishing boundary conditions 
\begin{equation}
\phi_n(0)=0 \hspace{1cm}\text{and}\hspace{1cm} \phi_n(\infty)=0\;.
\end{equation}
The authors of Ref.~\cite{Karch:2006pv} showed that, by doing the substitution
\begin{equation}
\label{eq:substitution}
\phi_n(z) = \e^{\kappa^2 z^2/2}\,\sqrt{z}\, \Psi_n(z)\;,
\end{equation}
one transforms the equation (\ref{eq:eqmotionphi}) into a Schr\"{o}dinger equation for $\Psi_n(z)$
\begin{equation}
\label{eq:schroedinger}
-\Psi_n^{\prime\prime} + \left(\kappa^4 z^2 + \frac{3}{4z^2}\right)\Psi_n = M_V(n)^2 \, \Psi_n \;,
\end{equation}
which is exactly solvable in terms of generalized Laguerre polynomials ${\rm L}^{(\alpha)}_{n}$.  The solutions $\phi_n(z)$ of the original equation (\ref{eq:eqmotionphi}) are then
\begin{align}
\label{eq:KK}
\phi_n(z) =\kappa^2 z^2\sqrt{\frac{2}{n+1}} \, {\rm L}^{(1)}_{n}(\kappa^2 z^2) \;,
\end{align}
and the required normalization conditions are given for each representations by 
\begin{equation}
\int_0^\infty \d z \; w_0(z) \, \phi_m(z)\, \phi_n (z) = \delta_{mn} \hspace*{1cm} \text{and} \hspace*{1cm}   \int_0^\infty \d z  \,  \Psi_m(z)\, \Psi_n (z) = \delta_{mn}\;.
\end{equation}

The most important result, which characterizes the Soft-Wall model and makes it a possible improvement, with respect to the HW models, concerns the resulting  mass spectrum which was found in \cite{Karch:2006pv} to be 
\begin{equation}
\label{eq:massKarch}
M_V(n)^2 = 4\kappa^2(n+1)
\end{equation}
with $ n = 0,1, \dots $. Eq. (\ref{eq:massKarch}) shows that the infinite set of 4D resonances corresponding to the 5D eigenfunctions {\it follows a Regge trajectory}. 

Despite this very nice approach to understand the Regge trajectory, the  SW model does not reproduce the correct OPE of the vector correlator. In fact,  the large-$Q^2$ expansion of two-point function coming from the solution of (\ref{eq:eqmotionvector}) has a non vanishing cofficient for the $\sfrac{1}{Q^2}$ term, corresponding  a non vanishing  dimension-two condensate, in contrast to what predicted in QCD (a fact that was already pointed out in \cite{Andreev:2006vy,Cata:2006ak,Zuo:2008re}). Also the values of  higher dimension condensates turn out to be different from those obtained in QCD.

The first purpose of this paper will be  build a deformation of the dilaton field such that the OPE of the vector correlator corresponds to the real one. 

\subsection{The vector two-point function}

One defines the vector-vector correlator  $\Pi_V$ ,
\begin{equation}
i \int \d^4x  \; \e^{ - i q \cdot x} \left\langle J_V^a\,_{\mu}(x)\; J_V^b\,_{\nu}(0) \right\rangle = \delta^{ab}\,\left(q_\mu q_ \nu - q^2 \eta_{\mu \nu} \right) \Pi_V(-q^2) \ .
\end{equation}
where $q$ is the 4D momentum and the current  $J_V^a\,_{\mu}$ is the one defined before equation (\ref{eq:vectorialsource}).

The function $f_V(-q^2,z)$ satisfying eq. (\ref{eq:eqmotionvector}), (\ref{eq:fv=1}) and vanishing at $z \rightarrow \infty$, the so-called the bulk-to-boundary propagator, plays a pivotal role in holographic models, since its knowledge is fundamental in  evaluating Green functions of the 4D theory. Once $f_V$ has been found, it can be substituted in the quadratic part of the 5D action (\ref{eq:lagragian5DVect}), which is the one relevant for the evaluation of the two-point function. The resulting 5D expression reduces to a 4D boundary term, quadratic in the 4D source $\mathsf{v}_\mu$, from which one gets the two-point function as the limit (with the Euclidean momentum $Q^2 \doteq -q^2$)
\begin{equation}
\label{eq:PiVfv}
Q^2 \Pi_V(Q^2) = \frac{1}{g_5^2} \lim_{z\rightarrow 0} w_0(z) f_V(Q^2,z) \; \partial_{z} f_V(Q^2,z)\;.
\end{equation}

Our main concern is that the two-point function $\Pi_V$ coming from the model built in 5D has the following properties: 
\begin{enumerate} [\itshape (i)]

\item \textbf{The progression of its poles in the Minkowski region (the resonances) follows a Regge trajectory.} 

In the Large-N$_c$ limit of QCD, the two-point function $\Pi_V$ could be written as a sum over an infinite set of stable vector resonances \cite{thooft,witten}
\begin{equation}
\label{eq:DefLargeNcPiV}
\Pi_V(Q^2) = \sum_{n=0}^\infty \frac{F_V(n)^2}{Q^2+M_V(n)^2}\;,
\end{equation}
where $F_V(n)$ are named decay constants and the $M_V(n)$ are the masses associated to the resonances of the vectorial channel: $\rho$, $\rho^\prime$, $\rho^{\prime\prime}$,... . We assume that these resonances follow, in a first approximation, a Regge progression \cite{collins},  
\begin{equation}
\label{eq:regge}
M_V (n)^2 \underset{n\rightarrow \infty}{\thicksim} \sigma n  \;,
\end{equation}  
where the integer $n$ is the radial excitation number and $\sigma$ is related to the confining string tension as explained in \cite{Karch:2006pv} and we can evaluate $\sigma \approx 0.90$ GeV$^2$ from \cite{Beringer:1900zz} as illustrated in figure \ref{fig:regge}. As pointed out in \cite{Masjuan:2012gc} by their authors, one has to organize the rho resonances according to their radial and their angular-momentum  progressions in order to select the right Regge trajectory. It means that we should not consider all set of resonances in figure (\ref{fig:regge} a) but only the ones in figure (\ref{fig:regge} b) corresponding to the masses squared progression   $M_V(n)^2 \sim 1.43(13) n $ GeV$^2$ or $\sigma \approx 1.43(13)$ GeV$^2$ \footnote{We thank Pere Masjuan to have pointed out this observation.}.
\begin{figure}[h]
\begin{minipage}[c]{.46\linewidth}

\begin{tikzpicture}[scale=0.6,inner sep=2mm]
\begin{axis}[xlabel={$n$},ylabel={$M_V(n)^2$ in GeV$^2$},ymin=0,xmin=0]
\addplot+[no marks,blue,dashed,domain=0:6.6] {0.9*x};
    \addplot[nodes near coords, only marks,mark=square*,red,fill=red, point meta=explicit symbolic]
    coordinates {
        (1,0.591361) [$\black \rho(770)$]
        (2,2.14623) [$\black \rho(1450)$]
        (3,2.9584) [$\black \rho(1700)$]
        (4,3.61) [$\black \rho(1900)$]
        (5,4.6225) [$\black \rho(2150)$]
        (6,5.1529) [$\black \rho(2270)$]
    };
\end{axis}
\end{tikzpicture}
\begin{center}
\vspace*{-0.5cm}(a)
\end{center}
 \end{minipage} \hfill
   \begin{minipage}[c]{.46\linewidth}
\begin{tikzpicture}[scale=0.6,inner sep=2mm]
\begin{axis}[xlabel={$n$},ylabel={$M_V(n)^2$ in GeV$^2$},ymin=0,xmin=0]
\addplot+[no marks,blue,dashed,domain=0:4.5] {1.43*x-1.09};
    \addplot[nodes near coords, only marks,mark=square*,red,fill=red, point meta=explicit symbolic]
    coordinates {
        (1,0.591361) [$\black \rho(770)$]
        (2,2.14623) [$\black \rho(1450)$]
        (3,3.61) [$\black \rho(1900)$]
        (4,4.6225) [$\black \rho(2150)$]
    };
\end{axis}
\end{tikzpicture}
\begin{center}
\vspace*{-0.5cm}(b)
\end{center}
\end{minipage}   
\caption{The progression of the squared masses of the $\rho$ resonances according to their label $n$. The straight line represents the fitted relation. The (a) plot follows \cite{Karch:2006pv} for $M_V(n)^2 \sim 0.90 n$ GeV$^2$. The (b) plot represents the fit obtained in \cite{Masjuan:2012gc} with $M_V(n)^2 \sim 1.43(13) n$ GeV$^2$.}\label{fig:regge}

\end{figure}
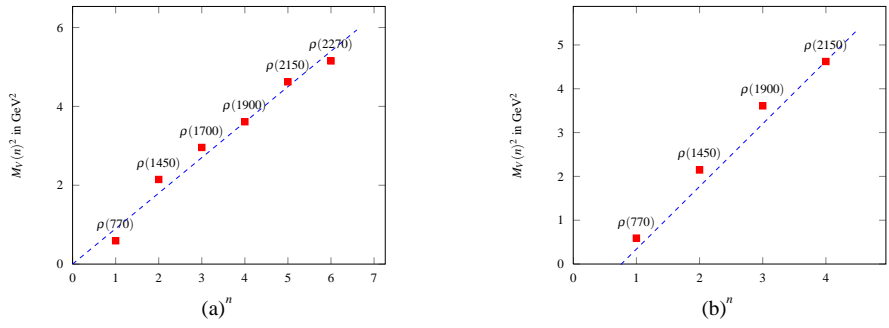
One assumes here that the Regge behaviour is more and more valid the larger $n$ is, which, of course, implies that the description of the lower resonances contributions would be less precise. From now on this spectrum will be called the physical spectrum.

\item \textbf{$\boldsymbol{\Pi_V}$ has the correct OPE.} 

One of the very well-known properties of this two point function is its OPE \cite{SVZ},  
\begin{equation}
\label{eq:OPEVect}
\Pi_V(Q^2) \underset{Q^2 \rightarrow \infty}{\thicksim}  \frac{1}{2}\frac{N_c}{12\pi^2}\ln \left(\frac{\Lambda_V^2}{Q^2}\right) + \left\langle \mathcal{O}_2\right\rangle\frac{1}{Q^2} + \left\langle \mathcal{O}_4\right\rangle \frac{1}{Q^4} + \left\langle \mathcal{O}_6\right\rangle_V \frac{1}{Q^6}
\end{equation}
where in the large-N$_c$ limit the coefficients of the OPE are given by

\begin{equation}
\label{eq:defOPEcoef}
\begin{cases}
&\left\langle \mathcal{O}_2\right\rangle  = 0  \\
&\left\langle \mathcal{O}_4\right\rangle  = \frac{1}{2} \frac{1}{12\pi} \;\alpha_s\; \left\langle G^2\right\rangle  \\
&\left\langle \mathcal{O}_6\right\rangle_V= \frac{1}{2}\left(-\frac{28\pi}{9}\right)  \;\alpha_s\; \left\langle \bar{\psi}\psi\right\rangle^2
\end{cases}\;,
\end{equation}
here $\alpha_s$ stands for the strong coupling constant, $\left\langle G^2\right\rangle$ for the gluon condensate, and $\left\langle \bar{\psi}\psi\right\rangle$ for the quark condensate.
\end{enumerate}

 The requirement that the two-point function obtained from the SW model (\ref{eq:lagragian5DVect}) with a suitable deformation of the dilaton profile, has the correct OPE (\ref{eq:OPEVect}) and (\ref{eq:defOPEcoef}), is the starting point of our analysis. Proposals to modify the original SW model based on the assumption of different profiles for the dilaton and for the vacuum expectation of a scalar field in the bulk were done in \cite{Kwee:2007nq,Gherghetta:2009ac}. 

\subsection{Vectorial OPE from a modified dilatonic background}
\label{sec:VectOPEMDB}

We shall assume that the effects of the OPE on the vector current two-point functions in QCD can be encoded in a new profile for the dilaton field of the SW model. More explicitly, we assume that OPE is related to the behaviour of the dilaton profile around the UV boundary, $z=0$ of the 5D, while keeping the metric to have the AdS form (\ref{eq:ds2}).  Moreover, as we want to keep the Regge behaviour induced by the dilaton profile $\Phi(z)$, we are led to modify the original quadratic profile of the SW dilaton by adding new terms which we collect in a function $B$,
\begin{equation}
\Phi(z) \longmapsto \Phi(z) + B(z) \;.
\end{equation}

We assume that the function $B$ can be represented \textit{for all} $z$ by a polynomial of degree $2K$ (with no constant term), 
\begin{equation}
\label{eq:defB}
B(z) = \sum_{k=1}^K \frac{b_{2k}}{2k} z^{2k}\;.
\end{equation}

Even for a polynomial form of the dilaton profile as given in (1) , the task of solving the corresponding equation of motion of the vector field is a non trivial one. 
In the following, we show how  the  \textit{ad hoc} peculiar form for $B(z)$ allows us to solve the equation of motion analytically, through an iterative method, and obtain the corresponding expression for the vector two-point function. 
\label{sec:DiffVEct}
To simplify the writing we define, 
\begin{equation}
w_0(z) \; \e^{-B(z)} \doteq w(z) \;.
\end{equation}

For counting the different orders of perturbation of the dilaton, we introduce an artificial control  parameter $\theta$ such as
\begin{equation}
\label{eq:wdeftheta}
w(z)=w_0(z) \e^{-B(\sqrt{\theta} z)}\;.
\end{equation}
We shall expand all relevant expressions in (formal) power series in $\theta$, using this parameter to translate the task of solving the differential equation for the vector field in the deformed dilatonic background in a more tractable set of iterative equations. 

At the end of the calculation,  we shall match the  asymptotic expansion of the two-point function to  the OPE expansion of QCD. Doing this the  dependence of the solution on the parameter $\theta$ disappears. We shall see that each  term $z^{2k}$ in the expression of $B(z)$ drives to a $\sfrac{1}{Q^{2k}}$ term in the OPE of the two point function and since we want to match the OPE only up to $\sfrac{1}{Q^{6}}$ term, we shall need only a polynomial of degree $6$, \textit{i.e.} to take  $K=3$ in (\ref{eq:defB}). The matching with the QCD OPE uniquely determines the coefficients $b_k$ of the dilaton profile. 

Let notice that, of course, other forms for $B(z)$ could give the same OPE but different exact solutions of the equation of motion to be interpreted as analytic continuation of the OPE. Our expression for $B$ is the simplest way to satisfy our assumptions \footnote{One could wonder if a complete knowledge of the full OPE expansion would allow us to write $B(z)$ as an infinite power series. Even in such an ideal case, that series would probably be only asymptotic and additional dynamical assumptions on how $B$ extends at any value of $z$ would be unavoidable, in the same manner that we have to deal with renormalons prescriptions \cite{Beneke:1998ui}.}.  

In the next Section \ref{sec:DiffVEct} we explain the technical details of the iterative method we derived to obtain the vector two-point function for the modified dilaton background. The complete expression for the two-point function is in \ref{app:VV}.

\subsection{Iterative construction of the vector two-point function for the modified dilaton background}
\label{sec:DiffVEct}

As we said before,  if we want to reproduce only up to the $\sfrac{1}{Q^6}$ term, the calculation has to be done up to order $\theta^3$.

Using (\ref{eq:wdeftheta}), the equation of motion for the vectorial field $\mathbb{V}$ associated to $w_0$ when expressed in its Fourier transform $f_V$ becomes, 
\begin{equation}
\label{eq:eqmotionvectorw}
\partial_z^2 \, f_V + \partial_z \left[ \ln w(z)\right]  \partial_z \, f_V  - Q^2\, f_V = 0 \;,
\end{equation}  
and can be decomposed into a hierarchical system of differential equations order by order in $\theta$. 

We can expand $f_V$ in power of $\theta$, 
\begin{equation}
f_V = f_V^{(0)} + \theta f_V^{(1)} + \theta^2 f_V^{(2)} + \theta^3 f_V^{(3)}+\mathcal{O}(\theta^4)
\end{equation}
with the boundary conditions
\begin{equation}
\label{eq:BCV}
f_V^{(n)}(Q^2,0) = \delta_{0,n} \hspace*{1cm} \text{and} \hspace*{1cm} f_V^{(n)}(Q^2,\infty)=0 \;.
\end{equation}

Then, the associated two-point function (\ref{eq:PiVfv}) has an expansion in $\theta$ too,  
\begin{equation}
\label{eq:Pivnint}
\Pi_V = \Pi_V^{(0)} + \theta \Pi_V^{(1)} + \theta^ 2 \Pi_V^{(2)} + \theta^3 \Pi_V^{(3)}+\mathcal{O}(\theta^4)\;,
\end{equation}
where
\begin{equation}
\label{eq:PiVn}
Q^2 \Pi_V^{(k)}(Q^2) = \frac{1}{g_5^2}\lim_{z\rightarrow 0} w_0(z) f_V^{(0)}(Q^2,z) \partial_{z} f_V^{(k)}(Q^2,z)\;,
\end{equation}
thanks to the boundary conditions (\ref{eq:BCV}).

To simplify a little the writing, let us define the differential operator $\mathfrak{D}$ such as 
\begin{equation}
\label{eq:Dfrak}
\mathfrak{D} \varphi \doteq \partial_z^2 \varphi + \partial_z \left[ \ln w_0(z)\right]  \partial_z \varphi  - Q^2  \varphi 
\end{equation}
then the equation $\mathfrak{D}f_V^{(0)} = 0$ corresponds to the unperturbed equation of motion (\ref{eq:eqmotionvector}). Organising order by order in $\theta$ the differential equations, one obtains the recursive system, 
\begin{equation}
\mathfrak{D}f_V^{(n)} = \sum_{k=0}^{n-1} z^{2(n-k)-1} \; b_{2(n-k)} \; \partial_z\,f_V^{(k)} \label{eq:DfV(n)}\;,
\end{equation}

or explicitly, for the first three orders, 
\begin{align}
\mathfrak{D}f_V^{(0)} &= 0\\
\mathfrak{D}f_V^{(1)} &= z \; b_2 \; \partial_z f_V^{(0)} \label{eq:DfV(1)}\\
\mathfrak{D}f_V^{(2)} &= z^3 \; b_4 \; \partial_z f_V^{(0)} + z \; b_2 \; \partial_z f_V^{(1)}	\\
\mathfrak{D}f_V^{(3)} &= z^5 \; b_6 \; \partial_z f_V^{(0)} + z^3 \; b_4 \; \partial_z f_V^{(1)} + z \; b_2 \; \partial_z f_V^{(2)}\label{eq:DfV(3)}\;.
\end{align}

To solve the system, we can use the Green function method. Indeed, from the first order correction (\ref{eq:DfV(1)}) on, one has the same differential equation with different source terms. To solve any equation with a source term $S$ like
\begin{equation}
\label{eq:eqdiffsource}
\mathfrak{D} f_V = S(Q^2,z) \;,
\end{equation}
one first solves the equation for the Green function $ G_V(Q^2;x,z)$,
\begin{equation}
\mathfrak{D} G_V(Q^2;x,z) = \delta(x-z)\;,
\end{equation}
with the boundaries conditions,
\begin{equation}
G_V(Q^2;0,z) = G_V(Q^2;\infty,z) = 0\;.
\end{equation} 

Then, the solution of (\ref{eq:eqdiffsource}) is given by 
\begin{equation}
f_V(Q^2,z) = \int_0^\infty \d x \;w_0(x)\; G_V(Q^2;z,x)\; S(Q^2,x)\;.
\end{equation}

Therefore, the differential system has for solution the recursive expression (for $n>0$), 
\begin{equation}
\label{eq:fVnint}
f_V^{(n)}(Q^2,z) = \int_0^\infty \d x \; w_0(x)\; G_V(Q^2;x,z) \left[\sum_{k=0}^{n-1} x^{2(n-k)-1} \; b_{2(n-k)} \; \partial_x\,f_V^{(k)}(Q^2,x)\right]\;,
\end{equation}
and using (\ref{eq:PiVn}), the solution for the vector two-point function is the convoluted expression, 
\begin{equation}
Q^2 \Pi_V^{(n)}(Q^2) =- \frac{1}{g_5^2} \sum_{k=0}^{n-1} b_{2(n-k)} \int_0^\infty \d x \; \e^{-\kappa^2x^2}\;  x^{2(n-k-1)} \; f_V^{(0)}(Q^2,x)  \; \partial_x\,f_V^{(k)}(Q^2,x) \;, 
\end{equation}
where we used the relation, 
\begin{equation}
f_V^{(0)}(Q^2,x) =- \lim_{z\rightarrow 0} w_0(z)f_V^{(0)}(Q^2,z) \partial_z G_V(Q^2;x,z)\;.
\end{equation}

In order to solve this system recursively, we have to know first $f_V^{(0)}$, which is nothing else that the well-known solution of the unperturbed equation of motion $\mathfrak{D}f_V^{(0)} = 0$ \textit{i.e.} the original bulk-to-boundary propagator of the SW model.  It could be expressed in several ways, for instance, as a series of the eigenfunctions, which in the SW model are given in terms of Laguerre polynomials (\ref{eq:KK}) \cite{Karch:2006pv} . 

However, in order to perform the calculation, we found more efficient to use the integral representation for $f_V^{(0)}$ given in  \cite{Grigoryan:2007my}, \textit{i.e.}  
\begin{equation}
\label{eq:f0vintegral}
f_V^{(0)}(Q^2,z) = \frac{Q^2}{4\kappa^2}\int_0^1 \d u \; u^{\frac{Q^2}{4\kappa^2}-1} \exp\left[- \frac{u}{1-u} \kappa^2 z^2 \right]\;;
\end{equation}
moreover, we constructed a similar representation also for the Green function (\textit{cf.}  \ref{app:fandG}), 
\begin{equation}
\label{eq:GVint}
G_V(Q^2;x,y)=-\frac{x y}{2} \int_0^1 \d t\,\; \frac{t^{\frac{Q^2}{4\kappa^ 2}-\frac{1}{2}}}{1-t} \exp \left[-\frac{t}{1-t}\kappa^2(x^2+y^2)\right]\mathrm{I}_1\left(2\kappa^2 x y \frac{\sqrt{t}}{1-t}\right)\;.
\end{equation} 
Leaving the details of the calculation to the  \ref{app:VV}, the final result for $\Pi_V^{(1)}$ is
\begin{equation}
\Pi_V^{(1)}(Q^2)=\frac{b_2}{4 \kappa^2 g_5^2 } \left(\frac{4 \kappa^2}{Q^2}\right) \left[ 1 +\left(\frac{Q^2}{4\kappa^2}\right) -  \left(\frac{Q^2}{4\kappa^2}\right)^2\psi^\prime\left(\frac{Q^2}{4\kappa^2}\right) \right]\;.
\end{equation}
In the same way, one can express order by order the analytic expression for $\Pi_V^{(n)}$ which takes the general form 
\begin{equation}
\label{eq:solvect}
Q^2 \Pi_V^{(n)}(Q^2) = \sum_{k=0}^n \mathcal{P}_{k,n}\left(\frac{Q^2}{4\kappa^2}\right) \; \psi^{(k)}\left(\frac{Q^2}{4\kappa^2}\right)\;,
\end{equation}
where, respectively, $\mathcal{P}_{k,n}$ are polynomials and $ \psi^{(k)}$ is the $k^\text{th}$ derivative of the Digamma $\psi$ function defined as the logarithmic derivative of the Euler's $\Gamma$ function. The coefficients of $\mathcal{P}_{k,n}$ depend only on $\kappa^2$ and the coefficients $b_k$ of the dilaton (\ref{eq:defB}). The explicit expressions for $\Pi^{(2)}$ and $\Pi^{(3)}$ are given respectively in eq. (\ref{eq:Pi2}) and (\ref{eq:Pi3}) in  \ref{app:VV}.

Our method provides a framework where we can get an explicit solution in terms of hypergeometric functions (see eqs. (\ref{eq:inttohyper}) and (\ref{eq:defhyper}) in  \ref{app:VV}), which then reduce to a combination of Digamma functions, and, importantly, it avoids the appearance of divergent series and divergent integrals, as it can be seen for instance, in the detailed calculation of $\Pi_V^{(1)}$ presented in  \ref{app:VV}. Moreover, as shown in  \ref{app:diagrams}, the use of such integral representations allows us to organize the calculation of each order correction in a very systematic way.

Let us comment here that the $\Pi_V^{(0)}$ function, 
\begin{equation}
Q^2 \Pi_V^{(0)}(Q^2)= -\frac{2 \kappa^2}{g_5^2} \left(\frac{Q^2}{4\kappa^2}\right)\left[ \gamma_E + \psi\left(\frac{Q^2}{4\kappa^2}+1\right)  \right]\;,
\end{equation}
can be associated to the \textit{ad hoc} models developed in \cite{Golterman:2001pj,Cata:2005zj,Jamin:2011vd} with the same Regge progression than us. From the Regge spectrum point of view itself, $\Pi_V^{(0)}$ is also similar to  expressions developed in \cite{deRafael:2012PJ} (see references therein) obtained by resummation over resonances. 

\subsection{Interpretation of the solution}

Given the complete analytic solution of $\Pi_V$, 
\begin{equation}
\label{eq:PiVntheta}
\Pi_V(Q^2) = \sum_{n=0}^3 \Pi_V^{(n)}(Q^2)\, \theta^n\;,
\end{equation}
we can ask ourself if \textit{(i) the poles of $\Pi_V$ in the Minkowski region still follow a Regge trajectory} and \textit{(ii) a correct OPE  is obtained}. We find that: 
\begin{enumerate} [\itshape (i)]
\item Since the part of each  $\Pi_V^{(n)}$ containing the Digamma function and its derivatives have poles only at all negative integers  $(-n)$,
\begin{equation}
\frac{-q^2}{4\kappa^2} = -n\;, 
\end{equation} 
at $0$-th order, only simple poles appear in $\Pi_V^{(0)}$, and so  the usual Regge spectrum (noticing also that the residue in $n=0$ is nought),
\begin{equation}
M(n)^2 = 4\kappa^2 n  = \sigma  n \;,
\end{equation}
is reproduced, schematically presented in figure \ref{fig:vectorialspectrum_2}. However, in eq. (\ref{eq:PiVntheta}), poles of higher order appear and a more detailed analysis has to be done to quantify possible departures from the original Regge spectrum.
\begin{figure}[ht]
\begin{center}
\begin{tikzpicture}[scale=0.8,inner sep=2mm]
\begin{axis}[hide y axis, tick style={color=black}, axis x line=bottom,xlabel=$n$,xmin=0,enlargelimits=true]
\addplot+[ycomb,red,fill=red,no marks,line width=2pt] plot coordinates {
        (1,2)
        (2,2)
        (3,2)
        (4,2)
    }; ;
\end{axis}
\end{tikzpicture}
\caption{ Vectorial Regge spectrum} \label{fig:vectorialspectrum_2}
\end{center}
\end{figure}
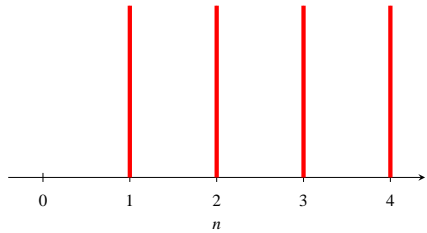

\item In order to reproduce the correct OPE up to the order $\sfrac{1}{Q^6}$, one can perform the asymptotic expansion for large $Q^2$ of $\Pi_V(Q^2)$ from the analytic expressions for the $\Pi_V^{(n)}$ in  \ref{app:VV}, 
\begin{align}
\Pi_V(Q^2) \underset{Q^2 \rightarrow \infty}{\thicksim}  & \frac{1}{2g_5^2} \ln \left(\frac{4 \kappa^2 \e^{-\gamma_E}}{Q^2}\right) + \frac{1}{2g_5^2} \left( 2 \kappa^ 2 + \theta b_2 \right)\frac{1}{Q^2} \nonumber\\
&+ \frac{1}{30g_5^{2}}\left[ -5 \left( 2 \kappa^2 + \theta b_2 \right)^2 +  20 \theta^2 b_4 \right] \frac{1}{Q^4} \nonumber\\
&+\frac{4}{5g_5^{2}} \left[- \left( 2 \kappa^2 + \theta b_2 \right) \theta^2 b_4 + 4 \theta^3 b_6)\right] \frac{1}{Q^6} \label{eq:OPEVectAnalytic}
\end{align}
and match (\ref{eq:OPEVectAnalytic}) term by term to the expected OPE 
\begin{equation}
\Pi_V(Q^2) \underset{Q^2 \rightarrow \infty}{\thicksim} \frac{N_c}{24\pi^2}\ln \left(\frac{\Lambda_V^2}{Q^2}\right) + \left\langle \mathcal{O}_2\right\rangle\frac{1}{Q^2} + \left\langle \mathcal{O}_4\right\rangle \frac{1}{Q^4} + \left\langle \mathcal{O}_6\right\rangle_V \frac{1}{Q^6}\;,
\end{equation}
with the values of the condensates given in (\ref{eq:defOPEcoef}). Notice that $\Lambda_V$ is automatically fixed to $2 \kappa  \e^{-\frac{\gamma_E}{2}} \approx 1$ GeV, which is the natural cut-off for our model. It implies also that
\begin{equation}
\label{eq:Fixingbs}
\begin{cases}
b_2  = - 2\kappa^2\\
\\
b_4 = \frac{3}{2} g_5^2\left< \mathcal{O}_4 \right>\\
\\
b_6 =  \frac{5}{16} g_5^2\left< \mathcal{O}_6 \right>_V \\
\end{cases}
\end{equation}
which completely fixes  $w(z)$ and hence the dilaton profile $B(z)$ in (\ref{eq:defB}). In (\ref{eq:Fixingbs}) we have removed the $\theta$ dependence since if one uses these expressions in $B(\sqrt{\theta} z)$, the artificial parameter $\theta$ disappears naturally. From now, we will no more make explicit the dependence in $\theta$ to simplify expressions for the reader except when this dependence is useful for the comprehension.
\end{enumerate}

Let us notice here, that as we anticipated in Sect. \ref{sec:VectOPEMDB}, (\ref{eq:Fixingbs}) shows that there is one-by-one correspondence between a $z^{2k}$ term in the function $B$ and a term at $\sfrac{1}{Q^{2k}}$ in the OPE. 

\subsection{Analysis of the vectorial spectrum}
\label{sec:vectspect}

From the complete and analytic expression for $\Pi_V$, one has shown that the poles of the spectrum follow a Regge behaviour, with masses squared $M(n)^2 = 4 \kappa^2 n$. The Large-N$_c$ representation of $\Pi_V$ in (\ref{eq:DefLargeNcPiV}) implies that the residues associated to the masses poles define the so-called Large-N$_c$ decay constants
\begin{equation}
F_V(n)^2 \doteq  \mathsf{Res}\left[\Pi_V\left(Q^2 \right) , - 4 \kappa^2 n \right]\;, 
\end{equation}
with $F_V(0)=0$. They are obtained from the analytic expressions of each $\Pi^{(k)}$ given in \ref{app:VV}. Despite this canonical description of the vectorial correlator, the $F_V(n)^2$ cannot fully reconstruct  the analytic expression of $\Pi_V$. For example, let us focus on the order $\theta^2$ associated to the function $\Pi_V^{(2)}$. As expressed in (\ref{eq:Pi2}), 
\begin{align}
\Pi_V^{(2)}(Q^2)&=\frac{b_4}{\kappa^4g_5^2}\left(\frac{4\kappa ^2}{Q^2}\right)  \bigg[-2 - \left(\frac{Q^2}{4 \kappa ^2}\right)\left(5+6\frac{Q^2}{4 \kappa ^2}\right) \nonumber \\
&\hspace*{4cm}+ 2 \left(\frac{Q^2}{4 \kappa ^2}\right)^2\left(1+3\frac{Q^2}{4 \kappa ^2}\right) \psi ^{\prime}\left(\frac{Q^2}{4 \kappa ^2}\right) \bigg] \nonumber \\
&\;\;+\frac{b_2^2}{16\kappa^4g_5^2}\left[ -1+2 \left(\frac{Q^2}{4\kappa^2}\right)\psi ^{\prime}\left(\frac{Q^2}{4\kappa^2}\right)+\left(\frac{Q^2}{4\kappa^2}\right)^2\psi ^{\prime\prime}\left(\frac{Q^2}{4\kappa^2}\right) \right]\;, \label{eq:PiV2Spect}
\end{align}
and the corresponding residue progression is 
\begin{equation}
\mathsf{Res}\left[\Pi_V^{(2)}\left(Q^2 \right) , - 4 \kappa^2 n \right] =  \frac{b_4}{g_5^2}(1-6n)\;,
\end{equation}
that manifestly does not contain the part proportional to $b_2^2$ in (\ref{eq:PiV2Spect}). Then, whatever summation over resonances procedure is used, one cannot recover the complete function. More generally, the analytic expression (\ref{eq:solvect}) contains derivatives of the Digamma function, which, due to their expression
\begin{equation}
\psi^{(k)}\left(\frac{Q^2}{4\kappa^2}\right) = (-1)^{k+1}\Gamma(k+1)\sum_{n=0}^\infty\frac{1}{\left(\frac{Q^2}{4\kappa^2}+n\right)^{k+1}}\;,
\end{equation}
cannot be directly written as series of single poles. Physically, the previous expression of Digamma derivatives corresponds to the combination of $k+1$ propagators, this suggests that the analytic expression of the vectorial correlator is a combination of several propagators. If one looks at this problem from a "perturbation theory" point of view, it could be possible to \textit{re-encode} the contributions from several propagators into only one propagator but with a modified mass. In our case, it means that the higher order corrections in $\theta$, or equivalently in the OPE, could be expressed as subleading contribution to the Regge behaviour, as in Ref. \cite{Mondejar:2008dt},  for more details see \ref{app:KeepingRegge}.

\section{Axial two point function}
\label{sec:3}

In the previous section, we have presented a SW model with a deformed dilaton profile which describes Regge theory and allows us to obtain the correct OPE of the vectorial two-point function. The natural question is now if it is possible to do a similar construction for the axial correlator. The problem to recover the right axial spectrum and the right axial OPE is more complicated if we assume, as usual, that the coupling of the 5D vector and axial vector gauge field to the metric and the dilaton is the same. Actually, under these assumptions, the form of the dilaton profile is already fixed by the requirements of a correct OPE of the vector two-point function, with the coefficients $b_n$ identified to the vectorial OPE coefficients (\ref{eq:Fixingbs}). So we need new ingredients to reproduce the QCD patterns of chiral symmetry breaking, as we already mentioned at the beginning of Sect. \ref{sec:Preliminaries}.

Let first enlarge the gauge symmetry of the 5D Lagrangian to the whole group of chiral transformations $\mathrm{SU}(N_f)_L \times \mathrm{SU}(N_f)_R$, introducing gauge fields, denoted by $\mathbb{L}_M(x,z)$ and $\mathbb{R}_M(x,z)$  each transforming only under one of the two $\mathrm{SU}(N_f)$ groups. Furthermore, as in  \cite{Erlich:2005qh,Da Rold:2005zs}, we introduce also a 5D scalar field $\mathbb{X}(x,z)$ transforming under the bi-fundamental of the chiral group so that the gauge invariant action is taken as
\begin{multline}
S_5 = \frac{1}{2} \int \d^4x \int_0^{\infty} \d z \sqrt{-g}\,\e^{-\Phi(z)}\; \mathrm{Tr} \bigg\{ g^{MN}\left( D_{M} \mathbb{X}\right)^\dagger \left( D_{N} \mathbb{X}\right) - m^2 \left\vert \mathbb{X} \right\vert^2  \\
\left. -\frac{1}{4g_5^2} g_{MN}g_{RS}\left(\mathbb{F}_L^{MR}\mathbb{F}_L^{NS} +\mathbb{F}_R^{MR}\mathbb{F}_R^{NS}\right) \right\} \, ,
\end{multline}
where $ D^M \mathbb{X} = \partial^M  \mathbb{X} - i\, \mathbb{L}^M\, \mathbb{X}  + i\, \mathbb{X} \, \mathbb{R}^M$, $\mathbb{F}_L^{MN} = \partial^M \mathbb{L}^N \! - \partial^N \mathbb{L}^M \! - i  [\mathbb{L}^M \! , \mathbb{L}^N]$ and analogous expression for $\mathbb{F}_R^{MN}$.

The action, can be rewritten in terms of vector and axial fields, $\mathbb{V}=\sfrac{1}{2}( \mathbb{L} +  \mathbb{R})$ and $\mathbb{A}=\sfrac{1}{2}( \mathbb{L} -  \mathbb{R})$, and 
as shown in \cite{Erlich:2005qh,Da Rold:2005zs}, taking
\begin{equation}
\mathbb{X}^a \doteq \frac{v(z)}{2}\mathbb{I}^a, 
\end{equation}
the equation of motion for the Fourier transform over the 4D space of the axial field $\mathbb{A}$, $f_A(-q^2,z)$, becomes
\begin{equation}
\label{eq:eqmotionaxial}
\partial_z^2 \, f_A + \partial_z \left[ \ln w(z)\right]  \partial_z \, f_A  - Q^2 f_A  = g_5^2 \left(\frac{v(z)}{z}\right)^2 f_A\;,
\end{equation} 
while the equation of motion for the vector field remains unchanged. 

In this approach, chiral symmetry is broken by the 5D scalar field, and in particular by a non vanishing  profile $v(z)$. The form of $v(z)$ near the origin $z\sim 0$ is dictated in the AdS/CFT correspondence by asking  the field $\mathbb{X}$ to  be dual to the bilinear quark   $q\,\bar q$ operator, whose non vanishing VEV is responsible for spontaneous $\chi$SB in QCD. In the following, we shall assume for the dilaton profile $B(z)$ the one obtained from the OPE coefficients of the vectorial fields and then use a suitable form of the profile $v(z)$ to  encode the properties of the axial sector:  

\begin{itemize}
\item The axial spectrum contains a pion pole at $q^2=0$ and a Regge spectrum starting at $q^2 = M_{a_1}^2$ with the same spacing than the vectorial spectrum.
\item The axial OPE has the following expression 
\begin{equation}
\label{eq:OPEAxial}
\Pi_A(Q^2) \underset{Q^2 \rightarrow \infty}{\thicksim}\frac{N_c}{24\pi^2}\ln \left(\frac{\Lambda_A^2}{Q^2}\right) + \left\langle \mathcal{O}_2\right\rangle \frac{1}{Q^2} + \left\langle \mathcal{O}_4\right\rangle \frac{1}{Q^4} + \left\langle \mathcal{O}_6\right\rangle_A \frac{1}{Q^6}
\end{equation} and 
\begin{equation}
\label{eq:O6AO6V}
\left\langle \mathcal{O}_6\right\rangle_A = - \frac{11}{7}\left\langle \mathcal{O}_6\right\rangle_V\;.
\end{equation}
with the same definitions than in (\ref{eq:defOPEcoef}). 
\end{itemize}

\subsection{Realization of the axial spectrum}
\begin{figure}[h]
\begin{center}
\begin{tikzpicture}[scale=0.8,inner sep=2mm]
\begin{axis}[hide y axis, tick style={color=black}, axis x line=bottom,xlabel=$n$,xmin=0,enlargelimits=true]
\addplot+[ycomb,blue,fill=blue,no marks,line width=2pt] plot coordinates {
        (0,2)
        (2,2)
        (3,2)
        (4,2)
    }; ;
\end{axis}
\end{tikzpicture}

\caption{ Axial Regge spectrum} \label{fig:axialspectrum_3}
\end{center}
\end{figure}
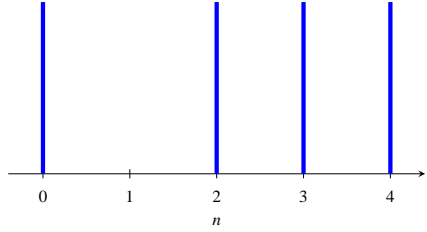

In order to satisfy the axial spectral properties, we make the following first ansatz for the contribution due to the scalar profile in (\ref{eq:eqmotionaxial}):
\begin{equation}
\label{eq:profile_V1}
\left(\frac{v(z)}{z}\right)^2 = \beta_{0} +\beta^* z\,\delta(z)\;.
\end{equation}

Let focus first on the Regge progression of the axial spectrum. We notice that, phenomenologically, taking  the first axial pole at $q^2 = M_{a_1}^2 \simeq  2 M_\rho^2 \simeq 2 \sigma$ is    quite an acceptable approximation for the axial spectrum. Then our prescription to obtain the axial spectrum from the vectorial one is to perform a "shift" over the vectorial spectrum like $Q^2\rightarrow Q^2 + 4\kappa^2$, as suggested from the comparison of the vectorial and axial mass spectra (see figures \ref{fig:vectorialspectrum_2} and \ref{fig:axialspectrum_3}). This fixes directly in (\ref{eq:eqmotionaxial})
\begin{equation}
g_5^2 \beta_0 = 4\kappa^2 = \sigma \;.
\end{equation}

The full axial spectrum is not yet realised, indeed, since the axial spectrum is a shifted version of the vectorial spectrum, it does not contain yet any pion pole and moreover on the Euclidean side, this shift implies a modifcation of the OPE as 
\begin{equation}
\Pi_A(Q^2) \underset{Q^2 \rightarrow \infty}{\thicksim} \frac{1}{2g_5^2}\ln \left(\frac{4\kappa^2\e^{-\gamma_E}}{Q^2}\right) +  \frac{2\kappa^2}{g_5^2} \frac{1}{Q^2} \;,
\end{equation}
where absence of dimension two operator in the axial OPE is violated.

The introduction of the Dirac delta function in (\ref{eq:profile_V1}) has the nice properties to cure this two problems at the same time. This term in (\ref{eq:eqmotionaxial}) generates only one exact contribution $-\sfrac{\beta^*}{Q^2}$ (for any $Q^2$) such that the pion pole appears naturally and modifying the OPE as 
\begin{equation}
\Pi_A(Q^2) \underset{Q^2 \rightarrow \infty}{\thicksim} \frac{1}{2g_5^2}\ln \left(\frac{4\kappa^2\e^{-\gamma_E}}{Q^2}\right) +\left( \frac{2\kappa^2}{g_5^2}-\beta^*\right) \frac{1}{Q^2}  \;,
\end{equation}
then by taking
 \begin{equation}
\beta^*=\frac{2\kappa^2}{g_5^2}\, ,\label{betastarvalue}
\end{equation}
the axial OPE properties is satisfied again. The existence of a  pion wave function associated  to this mechanism and its use in the evaluation of low energy chiral parameters will be exposed in details in a  forthcoming paper \cite{CDAG02}.

\subsection{Realization of the axial OPE}

The generation of the rest of the axial OPE terms derives exactly from the same procedure used in the vectorial section provided that we add two other terms in the expression (\ref{eq:profile_V1}) such as
\begin{equation}
\label{eq:profile_V2}
\left(\frac{v(z)}{z}\right)^2 = \beta_{0} +\beta^* z\,\delta(z) + \beta_2 z^2 + \beta_4 z^4 \;.
\end{equation}

With slight modifications, the iterative method we used to obtain the corrections to the vector two-point function, can be applied to the axial case too. All the details can be found in  \ref{app:AA}. Then one has by identification:  $\Lambda_A^2= 4\kappa^2\e^{-\gamma_E} = \Lambda_V^2$, the coefficients $\beta_2$ and $\beta_4$ are fixed by matching with QCD axial OPE (\ref{eq:OPEAxial}) ,(\ref{eq:O6AO6V}) :
\begin{equation}
\begin{cases}
\beta_2 & = -\frac{6\kappa^4}{g_5^2}\\
&\\
\beta_4 & = -\frac{10\kappa ^2}{3g_5^2}-5\kappa ^2\left\langle\mathcal{O}_4\right\rangle+\frac{45}{28}\left\langle\mathcal{O}_6\right\rangle_V\;.
\end{cases}
\end{equation}

The polynomial part of the scalar profile $v(z)$ is 
\begin{align}
v(z) & = z \sqrt{\beta_0 + \beta_2 z^2 + \beta_4 z^4} \nonumber\\
 & \underset{z\rightarrow 0}{\thicksim} \frac{2\kappa}{g_5}z - \frac{3\kappa^3}{2g_5}z^3 + \left(-\frac{67 \kappa ^5}{48g_5} -\frac{5g_5\kappa}{4}  \left\langle \mathcal{O}_4\right\rangle +\frac{45 g_5}{112 \kappa}\left\langle \mathcal{O}_6\right\rangle_V\right)z^5\;,
\end{align}
where the first two terms, which are the leading terms near $z=0$,  are exactly the ones of a scalar field dual to the bilinear quark $q\bar q$ operator, {\it i.e.}  the one required in Ref. \cite{Kwee:2007nq}  and compatible with the approach in \cite{Cox:2014zea}. 

The additional distributional term, proportional to $\beta^*$ in (\ref{eq:profile_V2}) can also be interpreted  as a boundary term, such that in the 5D action it would correspond to a peculiar 4D source term  in the generating functional, 
\begin{equation}
\label{eq:deltainS}
\beta^* \int \d^4 x \int_0^\infty \d z \; z \delta(z) \; w(z) \; \mathbb{A}_\mu(x,z)^2 = \beta^* \int \d^4 x \; \mathbb{A}_\mu(x,0)\; \mathbb{A}^\mu(x,0)\;.
\end{equation}

The appearance of the Dirac $\delta$ function term, dictated by the request of  correct resonance mass spectrum and OPE for the axial two-point function, makes our model definitely different from the more  usual approaches adopted for instance in \cite{Kwee:2007nq} and \cite{Colangelo:2011xk}.

\section{Analytic continuation in the chiral sector: the left-right correlator}
\label{sec:4}

In previous sections we have explicitly shown how to implement the constraints on an axial and vectorial two-point functions coming from two different regions in the $q^2-$plane: the deep Euclidean region where we reproduce the matching with the OPE of QCD, and the Minkowski region where the two-point function have poles following Regge trajectories. Having built explicit expressions for $\Pi_V$ and $\Pi_A$ valid on the whole complex plane we now turn to the analysis of their prediction for chiral quantities defined at low $Q^2$.

\subsection{The left-right correlator spectrum}

Since we are now interested to the chiral sector, \textit{i.e.} the low $Q^2$ expansions, it is more pertinent to consider the $\Pi_{LR}$ correlator,
\begin{equation}
\Pi_{LR}(Q^2) = \frac{1}{2} \left( \Pi_V(Q^2) - \Pi_A(Q^2) \right)\;.
\end{equation} 
which is an order parameter of the chiral symmetry breaking mechanism in QCD. 

\subsection{Predictions for the chiral constants}

The low $Q^2$ limit can be obtained by the analytic continuation of our expressions for the axial and vectorial two-point function, this allows us to extract from $\Pi_{LR}$ for instance the following chiral constants
\begin{align}
F_\pi^2 &=  2 \, \mathsf{Res}\left[\Pi_{LR}\left(Q^2 \right) , 0\right] \\
L_{10} & = \frac{1}{2} \frac{\d }{\d Q^2}\left[ Q^2 \Pi_{LR}(Q^2) \right] \bigg \vert_{Q^2=0}\;.
\end{align}
One has 
 
\begin{align}
F_\pi^2 & =   \beta^* + \left[ \frac{(\pi^2-9)\beta_2}{6\kappa^2}\right] \nonumber\\
&+ \left[\frac{90 \kappa ^2 \left(\beta_2 b_2 \left(4 \zeta (3)+5-\pi^2\right)-2 \left(\pi ^2-10\right) \beta_4\right)-45g_5^2 \beta_2^2 \left(-4
   \zeta (3)-5+\pi ^2\right)}{360 \kappa ^6}\right] \nonumber \\
&= \frac{N_c \kappa ^2 \left(180 \zeta (3)+191-41 \pi ^2\right)}{72 \pi^2}+\frac{5 \left(\pi ^2-10\right)}{2} \frac{\left\langle\mathcal{O}_4\right\rangle}{\kappa^2}-\frac{45 \left(\pi ^2-10\right)}{56}\frac{\left\langle\mathcal{O}_6\right\rangle_V}{\kappa ^4} \label{eq:FpiTheta}\;,
\end{align}
using $N_c=3$, $\kappa=\sqrt{1.43/4}\, \text{GeV}\simeq 0.6 \, \text{GeV} $ and the values of the condensates  $\left\langle \mathcal{O}_4\right\rangle = (- 0.635 \pm 0.04) \cdot 10^{-3}\,\text{GeV}^{4}$ and  $\left\langle \mathcal{O}_6\right\rangle_V= (14 \pm 3) \cdot 10^{-4}\,\text{GeV}^{6}$ from \cite{Golterman:2001nk}, we obtain
\begin{equation}
\label{eq:Fpinum}
F_\pi \simeq  \sqrt{4099.9 + 579 + 1147.8}\; \text{MeV} \simeq 76\; \left( \pm 3 \right)_{\text{ext.}} \;  \text{MeV}\;,
\end{equation}
the error in (\ref{eq:Fpinum}) are coming from the errors quoted for $\sigma$ and the condensates.  
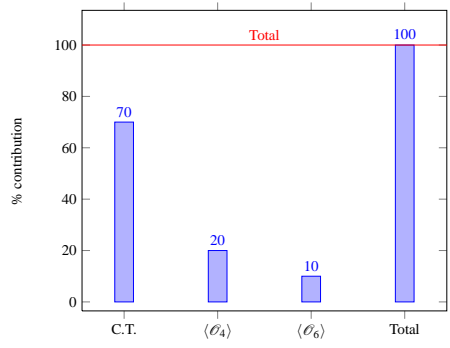
\begin{figure}[ht]
\begin{center}
\begin{tikzpicture}[scale=0.7]
\begin{axis}[ 
		ybar, 
		enlargelimits=0.15, 
		legend style={at={(0.5,-0.2)},anchor=north,legend columns=-1}, 
		ylabel={\% contribution}, 
		xticklabels={C.T., $\left\langle\mathcal{O}_4\right\rangle$, $\left\langle\mathcal{O}_6\right\rangle$, Total} ,
		xtick=data, 
		nodes near coords, 
		nodes near coords align={vertical}, ] 
		\addplot coordinates {(1,70) (2,20) (3,10) (4,100)}; 
		\addplot[red,sharp plot,update limits=false] coordinates {(0,100) (5,100)} node[above] at (axis cs:2.5,100) {Total};
\end{axis} 
\end{tikzpicture}
	\caption{Relative contributions to $F_\pi^2$ from eq. (\ref{eq:FpiTheta}). "C.T." stands for the contributions coming from the shift $g_5^2\beta_0$ and the contact term $g_5^2\beta^* z\delta(z)$. }\label{fig:FpiTheta} 
\end{center}
\end{figure}

In figure \ref{fig:FpiTheta}, the contributions to the final value of $F_\pi$ due to each term in (\ref{eq:FpiTheta}) are reported. 

The expression for $L_{10}$ is  
 
\begin{align}
L_{10} & = - \frac{1}{8g_5^2} +  \left[\frac{60 \left(\pi ^2-6\right)b_2 \kappa ^6+120g_5^2 \beta_2 \kappa^4 \left(-6 \zeta(3)-3+\pi ^2\right)}{5760 g_5^2 \kappa ^8}\right] \nonumber \\ 
&+\Bigg[\frac{-12 \pi ^4 \beta_2 b_2\kappa ^2-180 \pi ^2 \kappa ^2
   (3 \beta_4+\beta_2 b_2)+360 \beta_2 b_2\kappa ^2 (6 \zeta (3)+1)}{5760 \kappa ^8} \nonumber\\ 
   &\hspace*{1cm}+\frac{-30\kappa ^4 \left(-6 b_2^2 \zeta (3)-84 b_4+\pi ^2 \left(b_2^2+8 b_4\right)\right)+540 g_5^2 \beta_4 \kappa ^2 (4 \zeta (3)+5)}{5760 g_5^2\kappa ^8} \nonumber\\
   &\hspace*{6cm}+\frac{-6 g_5^2 \beta_2^2 \left(-30 (6 \zeta (3)+1)+15 \pi ^2+\pi ^4\right)}{5760\kappa ^8}\Bigg] \nonumber\\
&+\Bigg[\frac{\pi ^4 b_2^3 \kappa ^2+15 \pi ^2 \left(b_2^3+30 b_4b_2+72 b_6\right) \kappa ^2-180 \kappa ^2 \left(b_2^3 \zeta (3)+8 b_4b_2(\zeta (3)+2)+58 b_6\right)}{5760g_5^2 \kappa ^8}\Bigg]\nonumber\\
&=\frac{N_c (8010 \zeta (3)+495-585 \pi ^2-46 \pi ^4)}{8640\pi^2} \nonumber \\
&\hspace*{2cm}+\frac{-72 \zeta (3)-12+11 \pi ^2}{64}\frac{\left\langle\mathcal{O}_4\right\rangle}{\kappa^4}+\frac{5[5216 \zeta (3)+ 67 - 33 \pi ^2]}{1792}\frac{\left\langle\mathcal{O}_6\right\rangle_V}{\kappa ^6}\;,\label{eq:L10Theta}
\end{align}
then with the same numerical values used for the evaluation of $F_\pi$, 
\begin{equation}
\label{eq:L10num}
10^3 L_{10}\simeq -4.6 - 0.8 +0.1 \simeq -5.3 \left( \pm 1\right)_{\text{ext.}}\;,
\end{equation}
the error in (\ref{eq:L10num}) are coming from the errors quoted for  $\sigma$ and the condensates \footnote{Let us notice that if we had made the choice for $\sigma \simeq 0.9 \, \text{GeV}^2$ as in \cite{Karch:2006pv} the values obtained would have been $F_\pi \simeq 80$ MeV and $10^3 L_{10} \simeq -6.2$ which remain quite acceptable too.}. 

\begin{figure}[ht]
\begin{center}
 \begin{tikzpicture}[scale=0.7]
\begin{axis}[ 
		ybar, 
		enlargelimits=0.15, 
		legend style={at={(0.5,-0.2)},anchor=north,legend columns=-1}, 
		ylabel={\% contribution}, 
		xticklabels={C.T., $\left\langle\mathcal{O}_4\right\rangle$, $\left\langle\mathcal{O}_6\right\rangle$,Total} ,
		xtick=data, 
		nodes near coords, 
		nodes near coords align={vertical}, ] 
		\addplot coordinates {(1,87) (2,15) (3,-2) (4,100)}; 
		\addplot[red,sharp plot,update limits=false] coordinates {(0,100) (5,100)} node[above] at (axis cs:2.5,100) {Total};
\end{axis} 
\end{tikzpicture}
	\caption{Relative contributions to $L_{10}$ from eq. (\ref{eq:L10Theta}). "C.T." stands for the contributions coming from the shift $g_5^2\beta_0$ and the contact term $g_5^2\beta^* z\delta(z)$. }\label{fig:L10Theta}
\end{center}
\end{figure}
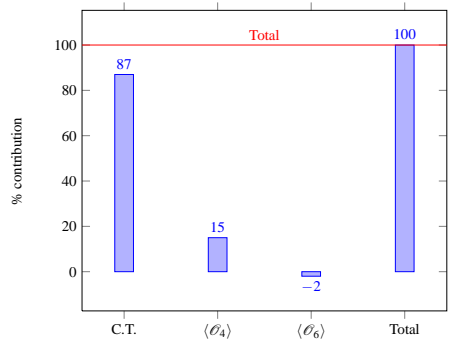
In figure \ref{fig:L10Theta}, the contributions to the final value of $L_{10}$ due to each term in (\ref{eq:L10Theta}) are reported.

\begin{figure}[h]
\begin{center}
\begin{tikzpicture}[scale=0.8]
\begin{axis}[xlabel={$\sigma$ in GeV$^2$},ylabel={$F_\pi$ in MeV},ymin=74,ymax=95,xmin=0.6,xmax=3]
\addplot[blue] table {FpiSigma.dat};
\end{axis}
\end{tikzpicture}
\caption{$F_\pi$ variation according to $\sigma$ } \label{fig:FpiSigma}
\end{center}
\end{figure}
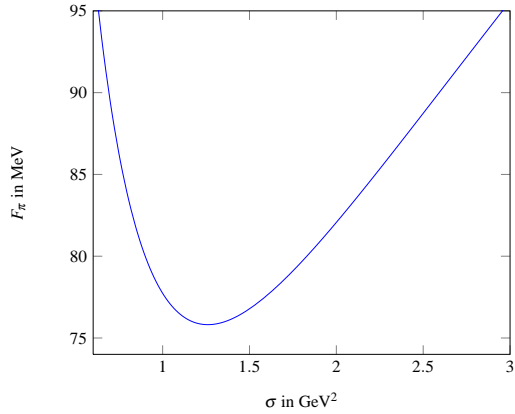

The values of $F_\pi$ and $L_{10}$ are quite reasonable face with our model hypothesis, if one admits the usual 30\% error coming from Large-N$_c$ QCD limit, compared to the range of variation of $F_\pi$ in the chiral limit, $66<F_\pi < 84$ MeV, and compared to the value $10^3L_{10} = -5.3 \pm 0.13$ according to \cite{Ecker:2013xja} and references therein. We can also notice here that the value of $F_\pi$ is strongly related to the value of $\kappa^2$ and  then to $\sigma=4\kappa^2$ as illustrated on figure \ref{fig:FpiSigma}.

The relative contributions to $F_\pi$ and $L_{10}$ from the gluon and the quark condensates are consistent with previous evaluation \cite{deRafael:1995zv,Hirn:2005vk}.

\subsection{Comments on the duality violation}

The so-called \textit{duality violation} phenomena \cite{shi94,shi95,cdsu96,bsz98,shi00} is a manifestation of the loss of the quark-hadron duality, \textit{i.e.}, the two descriptions of the same Green function of QCD, the one in the Minkowski region ($\R q^2 \geqslant 0$) as a collection (series) of bound states (resonances) and the one valid in the (deep) Euclidean region ($\R q^2 \leqslant 0$), are not related by analytic continuation. It implies that each expression is valid only in their own region. This two regions in the complex $q^2-$plane are disconnected and separated by the so-called Stokes' line. Until now, nobody knows where the QCD Stokes' line is, but one suspects that it could be the imaginary axis.    

Our model provides analytic expressions for the vectorial and axial correlators, it is natural to look for such duality violations, since it is easy to recover the different behaviours of the functions both in Minkowski and Euclidean regions. In our analysis a crucial role is played by the second Weinberg sum rule of $\Pi_{LR}$, whose importance in duality violations of QCD sum rules has been studied, also at a quantitative level in \cite{GonzalezAlonso:2010xf}.

As we already said in section \ref{sec:DiffVEct}, $\Pi_V^{(0)}$ corresponds to the model exposed by the authors of \cite{Cata:2005zj,Jamin:2011vd} and their discussion about the duality violation is still valid in our case, since our work is a "natural" extension including higher derivatives of the Digamma function $\psi$ in (\ref{eq:solvect}). 

The duality violation $\Delta = \sum_n \Delta^{(n)} \theta^n $ (developed in powers of $\theta$) is then
\begin{equation}
\Delta^{(n)} (q^2) = \Pi^{(n)}(\vert q^2\vert ) - \Pi^{(n)}(-\vert q^2\vert )\;,
\end{equation}
where the functions $\Pi^{(n)}$ are given in the  \ref{app:VV}.  $\Delta^{(n)}$ is not vanishing in our model, since,
\begin{equation}
\label{eq:psiminuspsiplus}
\psi^{(k)}(-a) - (-1)^k\psi^{(k)}(a) = \frac{\d^k}{\d a^k} \left[ \frac{1}{a} + \cot \pi a \right]\;,
\end{equation}  
where for simplicity, $a=-|q^2|/4\kappa^2$. Therefore our model has a natural implementation for duality violations that will be investigated in a future work. 

\section{Conclusions}
\label{sec:5}

There have been several models describing QCD properties of the  two point  vectorial and axial Green function. Already an interpolation among low energy chiral properties and perturbative QCD is good \cite{Peris:1998nj}. The intermediate region could be phenomenologically matched with a tower of resonance states equally spaced (\`a la Regge) \cite{Golterman:2001pj, Golterman:2001nk, Cata:2005zj, Shifman:2007xn}. Indeed, there are excellent dynamical reasons that Regge trajectories are dynamically generated. Holographic QCD gives us a fundamental theoretical tool with the SW model to start directly from a theory where Regge trajectories are analytically implemented. The original model \cite{Karch:2006pv} had several difficulties: the absence of a satisfying description of chiral symmetry breaking and of the axial sector, wrong OPE. For some further attempts to cure these problems one can refer to \cite{deTeramond:2009xk,Batell:2008zm}.

It is natural to modify opportunely the SW model to comply with the OPE. While it was done already for the HW model \cite{Cappiello:2009cj,Hirn:2005vk}, it is a novelty for the vector correlator in the SW model: we obtain the solution for the vector field $f_V$ profile and the analytic expression $\Pi _V ^{(n)} (Q^2)$ in eq. (\ref{eq:solvect}) in terms of the Digamma function $\psi$ and its derivatives, from the set of differential equations in (\ref{eq:DfV(n)})  ($n$ is related to the order of  OPE that we are addressing) --  \ref{app:fandG} and \ref{app:VV} are dedicated to a description of the calculation. As a result, we support previous works  \cite{ Golterman:2001pj,Cata:2005zj,shi00}.  

The coefficients of the polynomial of the dilaton profile ($b_k$'s), fixed by the requirement  of a correct  vectorial OPE,  appear also  in the  differential equation  (\ref{eq:eqmotionaxial}) of the axial vector field $f_A$, whose expression   extends the analogous ones in Ref. \cite{Son:2003et,Erlich:2005qh,Kwee:2007nq,Da Rold:2005zs} with the presence of a scalar field in the bulk with a non-trivial vacuum profile, $v(z)$. Several issues have to be understood in connection also  for the determination of the profile $v(z)$: chiral symmetry breaking, pion pole, the correct axial spectrum and correct OPE. A coherent, complete and phenomenologically consistent solution emerges: due the phenomenologically observed relation $8\kappa ^2 = M_{a1}^2$ the solution for the axial in eq. (\ref{eq:eqmotionaxial}) can be obtained from the vectorial one as $\Pi_A (Q^2)   =  \Pi_V  (Q^2 + 4\kappa ^2 ) + \text{"corrections"}$. This solution generates the pion pole
 , the correct axial spectrum and axial OPE, if the profile of the vacuum, $v(z)$, is not only a polynomial ($\beta_k$ in eq. (\ref{eq:EoMAxialwiththeta})) but contains also a Dirac $\delta$-function term, \textit{i.e.} a boundary term for the axial field,   which is needed in order to comply with the  axial OPE.

In  \ref{app:diagrams} we show an original pictorial representation of the solutions obtained, both for vectors and axial vectors, up to to the inclusion of the  term  $z^6$ in the polynomial part of the dilaton.

The necessity for simultaneous local and non-local (the Dirac $\delta$ function contribution) chiral symmetry breaking mechanism has been already noticed in the literature; for instance in Ref \cite{Domenech:2010aq} in the context of HW with  chiral symmetry broken by boundary conditions an additional scalar field with canonical AdS dimensions was added to describe the pion mass. In Ref \cite{Batell:2008zm} the behaviour $\Phi \sim z^2$ of the scalar field in the UV potentially has our same feature: the presence of an unwanted dimension-two condensate.

Our analytic solution for $F_\pi$ and $L_{10}$ respectively in eq. (\ref{eq:FpiTheta}) and (\ref{eq:L10Theta}) are very successfully phenomenologically and show how  these parameter are linked to Regge spacing and QCD condensates; compared to previous literature \cite{Golterman:2001pj, Golterman:2001nk, Cata:2005zj, Shifman:2007xn} our results are fully analytical.

We think there are many applications. For example, our model is the fitting framework to analyse the duality violations since they appear naturally as the analytic continuation properties of the Digamma function (and its derivatives) \textit{cf.} eq. (\ref{eq:psiminuspsiplus}). We enforce and extend here properties established from \textit{ad hoc} models in Ref. \cite{Cata:2005zj,Jamin:2011vd,shi00}; in our case these properties are based on a  (5D) Lagrangian description and analytic results for the solutions of the corresponding equations of motion.

\section*{Acknowledgements}

We thank Brian Batell, Martin Beneke, Marc Knecht, Pere Masjuan, Santiago Peris, Antonio Pineda, Eduardo de Rafael and Andreas Wulzer for discussions. G. D'A. and D.G. thank the organizers of the conference \textit{"Resurgence and Transseries in Quantum, Gauge and String Theories"} at CERN to have provided a nice opportunity for exchanges. G. D'A. is grateful to the Dipartimento di Fisica of Federico II University, Naples, for hospitality and support and acknowledges partial support by MIUR under project 2010YJ2NYW. L.C. and D.G.  were  supported in part  by MIUR under project 2010YJ2NYW  and by INFN research initiative PhenoLNF.

\appendix

\section{Integral representation of the Green function}
\label{app:fandG}

The solution of the unperturbed equation of motion $(\mathfrak{D}f_V^{(0)} = 0)$ could be expressed with generalized Laguerre polynomials ${\rm L}^{(\alpha)}_n$ or by an integral representation \cite{Karch:2006pv,Grigoryan:2007my}
\begin{equation}
\label{eq:fv0app}
f_V^{(0)}(Q^2,z) =  \kappa^2 z^ 2 \sum_{n=0}^\infty \frac{{\rm L}^{(1)}_n(\kappa^2z^2)}{\frac{Q^2}{4\kappa^2}+(n+1)} = \frac{Q^2}{4\kappa^2}\int_0^1 \d u \; u^{\frac{Q^2}{4\kappa^2}-1} \exp\left[- \frac{u}{1-u} \kappa^2 z^2 \right]\;.
\end{equation}

The associated Green function is given by
\begin{equation}
G_V(Q^2;x,y) =  - \frac{\kappa^2 x^2 y^2}{2}\sum_{n=0}^\infty \frac{{\rm L}^{(1)}_n(\kappa^2x^2)\,{\rm L}^{(1)}_n(\kappa^2y^2)}{n+1}\frac{1}{\frac{Q^2}{4\kappa^2}+(n+1)}\;,
\end{equation}
such that one has explicitly
\begin{equation}
f_V^{(0)}(Q^2,x) =- \lim_{z\rightarrow 0} w_0(z)f_V^{(0)}(Q^2,z) \partial_z G(Q^2;x,z)\;,
\end{equation}
since ${\rm L}^{(1)}_n(0)=n+1$.

It is more useful for our calculations to have an integral representation for the Green function too. One can express the Green function as an integral by using the Poisson Kernel \cite{Rainv:1960} 
\begin{multline}
\sum_{n=0}^\infty \frac{\Gamma(n+1)}{\Gamma(n+1+\alpha)} {\rm L}^{(\alpha)}_n(x) {\rm L}^{(\alpha)}_n(y) t^n \\
= (x y t)^{-\frac{\alpha}{2}} (1-t)^{-1} \exp\left[-\frac{t}{1-t}(x+y)\right] \mathrm{I}_\alpha\left(\frac{2\sqrt{x y t }}{1-t} \right) 
\end{multline}
where $\mathrm{I}_\alpha$ is the Bessel function, $|t|<1$ and $\alpha>-1$, and since $\alpha=1$ in our case, writing 
\begin{equation}
\frac{1}{\frac{Q^2}{4\kappa^2}+(n+1)}= \int_0^1\d t \;t^{\frac{Q^2}{4\kappa^ 2}} \; t^n\;,
\end{equation}
then 
\begin{align}
G_V(Q^2;x,y) &= -  \frac{\kappa^2 x^2 y^2}{2} \int_0^1 \d t\,\;t ^{\frac{Q^2}{4\kappa^ 2}} \sum_{n=0}^\infty \frac{\,{\rm L}^{(1)}_n(\kappa^2x^2)\,{\rm L}^{(1)}_n(\kappa^2y^2)}{n+1} t^n\\
&=-\frac{x y}{2} \int_0^1 \d t\,\; \frac{t^{\frac{Q^2}{4\kappa^ 2}-\frac{1}{2}}}{1-t} \exp \left[-\frac{t}{1-t}\kappa^2(x^2+y^2)\right]\mathrm{I}_1\left(2\kappa^2 x y \frac{\sqrt{t}}{1-t}\right) \label{eq:GVintapp}\;.
\end{align}

\section{Analytic expression for the vectorial two point function}
\label{app:VV}

The procedure that we use for the evaluation of the analytic expression for the two point function is quite simple. We first use the integral representations (\ref{eq:fv0app}) and (\ref{eq:GVintapp}), then we perform the integrations over the exponentials and Bessel's functions in the integrand and then there remain only integrals of rational functions. As an example let focus on $\Pi_V^{(1)}$ calculation, 
\begin{align}
\Pi_V^{(1)}(Q^2) &=-\frac{b_2}{g_5^2} \frac{1}{Q^2} \int_0^\infty \d x \;w_0(x) x \partial_x f_V^{(0)}(Q^2,x) \lim_{z \rightarrow 0} w_0(z) f_V^{(0)}(Q^2,z) G_V(Q^2;z,x) \label{eq:Pi(1)fG}\\
&= \frac{b_2}{g_5^2} \frac{1}{Q^2} \int_0^\infty \d x \; \e^{-\kappa^2 x^2}\;f_V^{(0)}(Q^2,x)\;\partial_x f_V^{(0)}(Q^2,x)\\
&= \frac{b_2}{g_5^2} \frac{1}{Q^2}  \left( \frac{Q^2}{4\kappa^2}\right)^2\int_0^\infty \d x \; \e^{-\kappa^2 x^2} \int_0^1 \d u \, \d v \;   u^{\frac{Q^2}{4\kappa^2}-1} v^{\frac{Q^2}{4\kappa^2}-1}\nonumber \\
&\hspace*{4cm}\times \left( -2 \kappa^2 x \frac{v}{1-v}\right) \exp\left[- \left(\frac{u}{1-u} +\frac{v}{1-v} \right) \kappa^2 x^2 \right] \label{eq:Pi1Intx}\\
&= \frac{b_2}{g_5^2}  \frac{1}{Q^2}  \left(\frac{Q^2}{4\kappa^2}\right)^2 \int_0^1 \d u \, \d v \;   u^{\frac{Q^2}{4\kappa^2}-1} v^{\frac{Q^2}{4\kappa^2}} \frac{(1-u)}{1- u v} \label{eq:Pi1intuv}\\
&=\frac{b_2}{4 \kappa^2 g_5^2 } \frac{1}{\left(1+\frac{Q^2}{4 \kappa^2}\right)^2} \hyper{3}{2}{1,\frac{Q^2}{4 \kappa^2},\frac{Q^2}{4 \kappa^2}+1}{2+\frac{Q^2}{4 \kappa^2},2+\frac{Q^2}{4 \kappa^2}}{1}\\
&=\frac{b_2}{4 \kappa^2 g_5^2 } \left(\frac{4 \kappa^2}{Q^2}\right) \left[ 1 +\left(\frac{Q^2}{4\kappa^2}\right) -  \left(\frac{Q^2}{4\kappa^2}\right)^2\psi^\prime\left(\frac{Q^2}{4\kappa^2}\right) \right] \\
&\underset{Q^2 \rightarrow \infty}{\thicksim} \frac{b_2}{8 \kappa^2 g_5^2 }\left[ \left(\frac{4\kappa^2}{Q^2}\right) - \frac{1}{3}\left(\frac{4\kappa^2}{Q^2}\right)^2 \right]\;.
\end{align} 

Each step can be applied to the evaluation of any higher order contributions. One always first performs the integrals over the fifth dimension ( $[0,\infty[$ ) as in equation (\ref{eq:Pi1Intx}), then one remains with multiple integrals between $[0,1]$ of the parameters entering in the integral representation of $f_V^{(0)}$ and $G_V$  -- cf (\ref{eq:fv0app}) and (\ref{eq:GVintapp})-- as in (\ref{eq:Pi1intuv}). For higher orders, this takes a general form which can be expressed as hypergeometric functions, 
\begin{multline}
\label{eq:inttohyper}
\int_0^1 \d u_1 \cdots \d u_n \; u^{a_1} (1-u_1)^{b_1}  \cdots   u^{a_n} (1-u_1)^{b_n}\frac{1}{(1-u_1\cdots u_n)^c} \\
= \Beta(1+a_1,1+b_1)\cdots\Beta(1+a_n,1+b_n)\hyper{n+1}{n}{c,1+a_1,\ldots,1+a_n}{2+a_1+b_1,\ldots,2+a_n+b_n}{1}\;,
\end{multline} 
where $\Beta$ is the Euler's beta function: $\Beta(x,y)=\frac{\Gamma(x)\Gamma(y)}{\Gamma(x+y)}$ and the hypergeometric function is  
\begin{equation}
\label{eq:defhyper}
\hyper{n+1}{n}{c,1+a_1,\ldots,1+a_n}{2+a_1+b_1,\ldots,2+a_n+b_n}{1} = \sum_{k=0}^\infty \frac{(c)_k (1+a_1)_k \cdots (1+a_n)_k}{(1)_k(2+a_1+b_1)_k\cdots(2+a_n+b_n)_k}\;,
\end{equation}
where the Pochhammer symbols are defined as $(a)_k = \Gamma(a+k)/\Gamma(a)$.
The hypergeometric functions $_p \mathrm{F}_q$ of  argument 1 have the particularity that their coefficients are function of $\frac{Q^2}{4\kappa^2} \pm 1$, so that in this case the series in (\ref{eq:defhyper}) can be summed in terms of Digamma functions and rational function \cite{Gotts:1988} as summarized in (\ref{eq:solvect}), \textit{viz.}
\begin{align}
\Pi_V^{(0)}(Q^2) &= \frac{1}{2g_5^2}\left[ \gamma_E + \psi\left(\frac{Q^2}{4\kappa^2}+1\right)  \right]\\
&\underset{Q^2 \rightarrow \infty}{\thicksim} \frac{1}{2g_5^2}\ln\left(\frac{Q^2}{4\kappa^2}\right) + \frac{1}{2g_5^2} \gamma_E  +\frac{1}{4g_5^2}\left(\frac{4\kappa^2}{Q^2}\right) - \frac{1}{24g_5^2} \left(\frac{4\kappa^2}{Q^2}\right)^2\nonumber \\
&\nonumber \\
\Pi_V^{(1)}(Q^2) &= \frac{b_2}{4 \kappa^2 g_5^2 } \frac{1}{\left(1+\frac{Q^2}{4 \kappa^2}\right)^2} \hyper{3}{2}{1,\frac{Q^2}{4 \kappa^2},\frac{Q^2}{4 \kappa^2}+1}{2+\frac{Q^2}{4 \kappa^2},2+\frac{Q^2}{4 \kappa^2}}{1}\nonumber \\
&= \frac{b_2}{4 \kappa^2 g_5^2 } \left(\frac{4 \kappa^2}{Q^2}\right) \left[ 1 +\left(\frac{Q^2}{4\kappa^2}\right) -  \left(\frac{Q^2}{4\kappa^2}\right)^2\psi^\prime\left(\frac{Q^2}{4\kappa^2}\right) \right] \label{eq:Pi1} \\
&\underset{Q^2 \rightarrow \infty}{\thicksim}\frac{b_2}{8 \kappa^2 g_5^2 }\left[ \left(\frac{4\kappa^2}{Q^2}\right) - \frac{1}{3}\left(\frac{4\kappa^2}{Q^2}\right)^2 \right] \nonumber\\
&\nonumber \\
\Pi_V^{(2)}(Q^2) &=\frac{b_4}{2\kappa^4g_5^2} \frac{1}{\left(1+\frac{Q^2}{4 \kappa^2}\right)^2\left(2+\frac{Q^2}{4 \kappa^2}\right)^2} \hyper{3}{2}{2,\frac{Q^2}{4 \kappa^2},\frac{Q^2}{4 \kappa^2}+1}{3+\frac{Q^2}{4\kappa^2},3+\frac{Q^2}{4 \kappa^2}}{1} \nonumber\\
&\;\;\;\;-\frac{b_2^2}{8\kappa^4g_5^2} \frac{1}{\left(1+\frac{Q^2}{4 \kappa^2}\right)^3\left(2+\frac{Q^2}{4 \kappa^2}\right)} \hyper{4}{3}{2,\frac{Q^2}{4 \kappa^2},\frac{Q^2}{4 \kappa^2}+1,\frac{Q^2}{4 \kappa^2}+1}{1+\frac{Q^2}{4 \kappa^2},2+\frac{Q^2}{4\kappa^2},3+\frac{Q^2}{4 \kappa^2}}{1}\nonumber\\
&=\frac{b_4}{\kappa^4g_5^2}\left(\frac{4\kappa ^2}{Q^2}\right) \left[-2 - \left(\frac{Q^2}{4 \kappa ^2}\right)\left(5+6\frac{Q^2}{4 \kappa ^2}\right) + 2 \left(\frac{Q^2}{4 \kappa ^2}\right)^2\left(1+3\frac{Q^2}{4 \kappa ^2}\right) \psi ^{\prime}\left(\frac{Q^2}{4 \kappa ^2}\right) \right] \nonumber\\
&\;\;\;+\frac{b_2^2}{16\kappa^4g_5^2}\left[ -1+2 \left(\frac{Q^2}{4\kappa^2}\right)\psi ^{\prime}\left(\frac{Q^2}{4\kappa^2}\right)+\left(\frac{Q^2}{4\kappa^2}\right)^2\psi ^{\prime\prime}\left(\frac{Q^2}{4\kappa^2}\right) \right] \label{eq:Pi2}\\
&\underset{Q^2 \rightarrow \infty}{\thicksim}  \frac{4b_4-b_2^2}{96\kappa^4g_5^2} \left(\frac{4\kappa^2}{Q^2}\right)^2 - \frac{b_4}{90\kappa^4g_5^2} \left(\frac{4\kappa^2}{Q^2}\right)^3 \nonumber \\ 
&\nonumber \\
\Pi_V^{(3)}(Q^2)&=\frac{6b_6}{\kappa^6g_5^2}\hyper{3}{2}{3,\frac{Q^2}{4 \kappa^2},\frac{Q^2}{4 \kappa^2}+1}{4+\frac{Q^2}{4 \kappa^2},4+\frac{Q^2}{4 \kappa^2}}{1} \nonumber\\
&\;\;\;-\frac{b_2b_4}{4\kappa^6g_5^2}\left(\frac{Q^2}{4\kappa^2}\right)\Bigg[ \frac{1}{\frac{Q^2}{4 \kappa^2}\left(1+\frac{Q^2}{4 \kappa^2}\right)^3\left(2+\frac{Q^2}{4 \kappa^2}\right)^2}\hyper{4}{3}{3,\frac{Q^2}{4 \kappa^2},\frac{Q^2}{4 \kappa^2}+1,\frac{Q^2}{4 \kappa^2}+1}{3+\frac{Q^2}{4 \kappa^2},3+\frac{Q^2}{4\kappa^2},2+\frac{Q^2}{4 \kappa^2}}{1} \nonumber \\
&-\frac{2}{\frac{Q^2}{4 \kappa^2}\left(1+\frac{Q^2}{4 \kappa^2}\right)\left(2+\frac{Q^2}{4 \kappa^2}\right)^3\left(3+\frac{Q^2}{4 \kappa^2}\right)}\hyper{4}{3}{3,\frac{Q^2}{4 \kappa^2},\frac{Q^2}{4 \kappa^2}+2,\frac{Q^2}{4 \kappa^2}+2}{3+\frac{Q^2}{4 \kappa^2},3+\frac{Q^2}{4\kappa^2},4+\frac{Q^2}{4 \kappa^2}}{1} \nonumber\\
&+\frac{1}{\left(1+\frac{Q^2}{4 \kappa^2}\right)\left(2+\frac{Q^2}{4 \kappa^2}\right)^3\left(3+\frac{Q^2}{4 \kappa^2}\right)^2}\hyper{4}{3}{3,\frac{Q^2}{4 \kappa^2}+1,\frac{Q^2}{4 \kappa^2}+2,\frac{Q^2}{4 \kappa^2}+2}{4+\frac{Q^2}{4 \kappa^2},4+\frac{Q^2}{4\kappa^2},3+\frac{Q^2}{4 \kappa^2}}{1}\nonumber\\
&+\frac{2}{\frac{Q^2}{4 \kappa^2}\left(1+\frac{Q^2}{4 \kappa^2}\right)^3\left(2+\frac{Q^2}{4 \kappa^2}\right)^2\left(3+\frac{Q^2}{4 \kappa^2}\right)}\hyper{4}{3}{3,\frac{Q^2}{4 \kappa^2},\frac{Q^2}{4 \kappa^2}+1,\frac{Q^2}{4 \kappa^2}+1}{2+\frac{Q^2}{4 \kappa^2},4+\frac{Q^2}{4\kappa^2},2+\frac{Q^2}{4 \kappa^2}}{1} \nonumber \\
&-\frac{2}{\left(1+\frac{Q^2}{4 \kappa^2}\right)\left(2+\frac{Q^2}{4 \kappa^2}\right)^3\left(3+\frac{Q^2}{4 \kappa^2}\right)\left(4+\frac{Q^2}{4 \kappa^2}\right)}\hyper{4}{3}{3,1+\frac{Q^2}{4 \kappa^2},2+\frac{Q^2}{4 \kappa^2},\frac{Q^2}{4 \kappa^2}+2}{3+\frac{Q^2}{4 \kappa^2},5+\frac{Q^2}{4\kappa^2},3+\frac{Q^2}{4 \kappa^2}}{1} \Bigg]\nonumber \\
&+\frac{b_2^3}{16\kappa^6g_5^2}\Bigg[\frac{1}{\frac{Q^2}{4 \kappa^2}\left(1+\frac{Q^2}{4 \kappa^2}\right)^4\left(2+\frac{Q^2}{4 \kappa^2}\right)}\hyper{5}{4}{3,\frac{Q^2}{4 \kappa^2},1+\frac{Q^2}{4 \kappa^2},1+\frac{Q^2}{4 \kappa^2},\frac{Q^2}{4 \kappa^2}+1}{3+\frac{Q^2}{4 \kappa^2},3+\frac{Q^2}{4\kappa^2},3+\frac{Q^2}{4 \kappa^2},3+\frac{Q^2}{4 \kappa^2} }{1} \nonumber\\
&-\frac{2}{\frac{Q^2}{4 \kappa^2}\left(1+\frac{Q^2}{4 \kappa^2}\right)^2\left(2+\frac{Q^2}{4 \kappa^2}\right)^2\left(3+\frac{Q^2}{4 \kappa^2}\right)}\hyper{3}{2}{3,\frac{Q^2}{4 \kappa^2},1+\frac{Q^2}{4 \kappa^2}}{3+\frac{Q^2}{4 \kappa^2},3+\frac{Q^2}{4\kappa^2},3+\frac{Q^2}{4 \kappa^2},3+\frac{Q^2}{4 \kappa^2} }{1}\nonumber\\
&+\frac{1}{\left(1+\frac{Q^2}{4 \kappa^2}\right)\left(2+\frac{Q^2}{4 \kappa^2}\right)^4\left(3+\frac{Q^2}{4 \kappa^2}\right)}\hyper{5}{4}{3,1+\frac{Q^2}{4 \kappa^2},2+\frac{Q^2}{4 \kappa^2},2+\frac{Q^2}{4 \kappa^2},\frac{Q^2}{4 \kappa^2}+2}{4+\frac{Q^2}{4 \kappa^2},4+\frac{Q^2}{4\kappa^2},4+\frac{Q^2}{4 \kappa^2},4+\frac{Q^2}{4 \kappa^2} }{1} \Bigg]\nonumber\\
&=\frac{b_6}{12\kappa^6g_5^2}\left(\frac{4 \kappa ^2}{Q^2}\right)\Bigg\{ 6 + \left(\frac{Q^2}{4 \kappa ^2}\right)\left[ 20 + 30 \left(\frac{Q^2}{4 \kappa ^2}\right)+ 33 \left(\frac{Q^2}{4 \kappa ^2}\right)^2 \right]\nonumber \\
&\hspace*{3.5cm} - 6 \left(\frac{Q^2}{4 \kappa ^2}\right)^2 \left[ 1 + 3 \left(\frac{Q^2}{4 \kappa ^2}\right) + 5 \left(\frac{Q^2}{4 \kappa ^2}\right)^2 \right] \psi^{\prime}\left(\frac{Q^2}{4 \kappa ^2}\right)\Bigg\} \nonumber\\
& -\frac{b_2b_4}{8\kappa^6g_5^2}  \left(\frac{4 \kappa ^2}{Q^2}\right)\Bigg\{ - 1 - \left(\frac{Q^2}{4 \kappa ^2}\right)\left[5+9\left(\frac{Q^2}{4 \kappa ^2}\right)\right]\nonumber\\
&\hspace*{3.5cm} +3\left(\frac{Q^2}{4 \kappa ^2}\right)^2\left[1+4\left(\frac{Q^2}{4 \kappa ^2}\right)\right] \psi ^{\prime}\left(\frac{Q^2}{4\kappa^2}\right) \nonumber\\
&\hspace*{3.5cm} + \left(\frac{Q^2}{4 \kappa ^2}\right)^3\left[1+\left(\frac{Q^2}{4 \kappa ^2}\right)\right] \psi ^{\prime \prime}\left(\frac{Q^2}{4\kappa^2}\right) \Bigg\}\nonumber\\
&-\frac{b_2^3}{16\kappa^6g_5^2} \Bigg\{-2 + 6 \left(\frac{Q^2}{4\kappa^2}\right) \psi ^{\prime}\left(\frac{Q^2}{4\kappa^2}\right) \nonumber \\
& \hspace*{4cm}+ 6 \left(\frac{Q^2}{4\kappa^2}\right)^2 \psi ^{\prime \prime}\left(\frac{Q^2}{4\kappa^2}\right)+ \left(\frac{Q^2}{4\kappa^2}\right)^3\psi^{\prime\prime\prime}\left(\frac{Q^2}{4\kappa^2}\right)\Bigg \} \label{eq:Pi3}\\
\underset{Q^2 \rightarrow \infty}{\thicksim} & \frac{ 4 b_6-b_2 b_4 }{80\kappa^6g_5^2} \left(\frac{4\kappa^2}{Q^2}\right)^3 \nonumber
\end{align}

\section{Analytic expression for the axial two point function}\label{app:AA}

We shall proceed in analogy with what we did in section \ref{sec:DiffVEct}. We expand,
\begin{equation}
f_A = f_A^{(0)} + \theta f_A^{(1)} + \theta^ 2 f_A^{(2)} + \theta^3 f_A^{(3)}+\mathcal{O}(\theta^4)
\end{equation}
and the associated axial two-point function,  
\begin{equation}
\Pi_A = \Pi_A^{(0)} + \theta \Pi_A^{(1)} + \theta^ 2 \Pi_A^{(2)} + \theta^3 \Pi_A^{(3)}+\mathcal{O}(\theta^4)\;.
\end{equation}
The expression (\ref{eq:profile_V2}) is also expanded in powers of $\theta$, 
\begin{equation}
\left(\frac{v(\sqrt{\theta} z )}{z}\right)^2 = \beta_0 + \beta_2 z^2 \theta + \beta_4 z^4 \theta^2 +\beta^* z\delta(z)\;.
\end{equation}
(The absence of powers of $\theta$ for the $\beta^*$ term is due to its  invariance under the rescaling $z\rightarrow \sqrt{\theta}z$.)

The equation of motion for the axial field (\ref{eq:eqmotionaxial}) becomes
\begin{equation}
\label{eq:EoMAxialwiththeta}
\partial_z^2 \, f_A + \partial_z \left[ \ln w(z)\right]  \partial_z \, f_A  - (Q^2+ g_5^2 \beta_0)f_A   =  g_5^2 \left[ \beta_2 z^2 \theta+ \beta_4 z^4 \theta^2+\beta^* z\delta(z) \right] f_A  \;, 
\end{equation}
with the boundary conditions
\begin{equation}
f_A^{(n)}(Q^2,0) = \delta_{0,n} \hspace*{1cm}\text{and}\hspace*{1cm}  f_A^{(n)}(Q^2,\infty)=0\;.
\end{equation}

The zero-th order solution of (\ref{eq:EoMAxialwiththeta}), neglecting the contribution from the $\beta^*$ term,  is given by (simply the implementation of the shift on the integral representation (\ref{eq:f0vintegral}) )
\begin{equation}
f_A^{(0)}(Q^2,z) = \left(\frac{Q^2}{4\kappa^2}+1\right)\int_0^1 \d x \; x^{\frac{Q^2}{4\kappa^2}} \exp\left[- \frac{x}{1-x} \kappa^2 z^2 \right]\;,
\end{equation}
and the associated Green function, 
\begin{equation}
G_A(Q^2;x,y)=-\frac{x y}{2} \int_0^1 \d t\,\; \frac{t^{\frac{Q^2}{4\kappa^ 2}+\frac{1}{2}}}{1-t} \exp \left[-\frac{t}{1-t}\kappa^2(x^2+y^2)\right]\mathrm{I}_1\left(2\kappa^2 x y \frac{\sqrt{t}}{1-t}\right)\;.
\end{equation}
The $\beta^*$ term contributes only to the $\theta^0$ term, and as it was the case for the vector correlator in Sect. \ref{sec:DiffVEct}, also for the axial one, we obtain, order by order in $\theta$:
\begin{multline}
\label{eq:PiArecsol}
Q^2 \Pi_A^{(n)}(Q^2)=-\beta^*\delta_{n,0}- \frac{1}{g_5^2}\int_0^\infty \d x \; \e^{-\kappa^2x^2}\; \; f_A^{(0)}(Q^2,x) \\
\times \sum_{k=0}^{n-1} \left[ g_5^2 \beta_{2(n-k)} x^{2(n-k)-1}\,f_A^{(k)}(Q^2,x)+ b_{2(n-k)} x^{2(n-k-1)}\; \partial_x\,f_A^{(k)}(Q^2,x)\right] \;.
\end{multline}

Using the same methodology one obtains the following contributions proportional to the $\beta_n$ coefficients in (\ref{eq:PiArecsol}) for the axial two point functions,
\begin{align}
\Pi_A^{(1)}(Q^2)\bigg \vert_\beta &= \frac{\beta_2}{8\kappa^4}\left(\frac{4\kappa^2}{Q^2}\right)^2 \left\{ 3 + 2\left(\frac{Q^2}{4\kappa^2}\right)^2 - 2 \left[1 + \left(\frac{Q^2}{4\kappa^2}\right) \right]^2 \psi^\prime\left(1+\frac{Q^2}{4\kappa^2}\right) \right\} \\
&\underset{Q^2 \rightarrow \infty}{\thicksim} \frac{\beta_2}{24\kappa^4} \left(\frac{4\kappa^2}{Q^2}\right)^2 + \frac{\beta_2}{24\kappa^4}\left(\frac{4\kappa^2}{Q^2}\right)^3 \nonumber 
\end{align}

\begin{align}
\Pi_A^{(2)}(Q^2)\bigg \vert_\beta &= \frac{\beta_4}{8\kappa^6} \left(\frac{4\kappa^2}{Q^2}\right)\Bigg\{ -10 - 3\left(\frac{Q^2}{4\kappa^2}\right)\left[ 5+2\left(\frac{Q^2}{4\kappa^2}\right)\right] \nonumber \\
&\hspace*{6cm}+6 \left[ 1+\left(\frac{Q^2}{4\kappa^2}\right)\right]^3 \psi^\prime\left(1+\frac{Q^2}{4\kappa^2}\right)  \Bigg\} \nonumber \\
&+\frac{g_5^2\beta_2^2+2\kappa^2\beta_2 b_2}{32\kappa^8}\left(\frac{4\kappa^2}{Q^2}\right)\Bigg\{ - 5 - 4\left(\frac{Q^2}{4\kappa^2}\right) +6 \left[1+\left(\frac{Q^2}{4\kappa^2}\right) \right]^2 \psi^{\prime\prime}\left(\frac{Q^2}{4\kappa^2} +1\right)\nonumber\\
&\hspace*{5cm}+2\left[1+\left(\frac{Q^2}{4\kappa^2}\right)\right]^3 \psi^{\prime} \left(\frac{Q^2}{4\kappa^2}\right)\Bigg\}\\
&\underset{Q^2 \rightarrow \infty}{\thicksim} -\frac{\beta_4}{40\kappa^6}\left(\frac{4\kappa^2}{Q^2}\right)^3 \nonumber
\end{align}

\section{Corrections to the mass spectrum}
\label{app:KeepingRegge}

In this Appendix, we want to show explicitly our assumption that Regge trajectories are kept in our approach and modified order by order in $\theta$ by sub-leading corrections in $n$. Indeed, the parameter $\theta$ introduced in Sect. \ref{sec:VectOPEMDB} acquires a physical meaning as a true perturbative parameter, when we try to quantify corrections to the leading Regge mass spectra. We develop the considerations we did at the end of Sect. \ref{sec:vectspect} for the vector resonance mass spectrum. The same argument would hold for the axial resonances as described in Sect. \ref{sec:4}.

As we showed in Sect. \ref{sec:DiffVEct},  the vector two-point function $\Pi_V(Q^2)$ receives corrections in $\theta$ eq. (\ref{eq:PiVntheta}) as 
\begin{equation}
\Pi_V(Q^2) = \sum_{k=0}^3 \Pi_V^{(k)}(Q^2)\, \theta^k\;,
\end{equation}
which generically take the form 
\begin{equation}
\label{eq:kterm}
\Pi_V^{(k)}(Q^2) = \sum_{n=1}^\infty \sum_{\ell=1}^{k+1}\frac{A^{(k)}_\ell(n)}{(Q^2 + 4\kappa^2 n)^\ell}\; ,
\end{equation}
showing the appearance of multiple poles of increasing order the greater the powers of $\theta$ considered in eq. (\ref{eq:solvect}).

The question arises whether it is possible to rewrite (\ref{eq:kterm})  in a way that preserves the Regge behaviour, maybe with the cost of modifying  the value of the slope parameter.  In fact, it is possible to answer in the affirmative, provided the value of $\theta$ is small enough.  It is possible to show that (\ref{eq:kterm}) can be generically rewritten such as 
\begin{equation}
\label{eq:PiVRegeeKeeping}
\Pi_V(Q^2) = \sum_{n=1}^\infty \frac{F_V(n,\theta)^2}{Q^2 + \sigma(n,\theta)\, n }\;,
\end{equation}
where
\begin{equation}
\sigma(n,\theta) = 4\kappa^2 + \frac{\sigma_{1}(\theta)}{n} + \cdots
\end{equation}
corresponding to corrections to the original Regge slope, $\sigma=4\kappa^2$, and to the residue. The equivalence between the two expressions (\ref{eq:kterm}) and (\ref{eq:PiVRegeeKeeping}) is non trivial, and relies on  a number of algebraic relations satisfied by the coefficients $A^{(k)}_\ell(n)$  that reduce the number of independent parameters thus  allowing the equivalent representation (\ref{eq:PiVRegeeKeeping}). Thus, for a given resonance term, {\emph{i.e.}} a fixed value of $n$, and a given value of the power of $\theta$ considered, the  departure from the original Regge slope can be made parametrically small by choosing a correspondingly small enough value of $\theta$. Notice that  coefficients of  higher power in $\theta$ are expected to be sizeable or even greater than the lower ones, following, at most, an asymptotic series behaviour.

In this context, the knowledge of the complete OPE series would drive to a complete resummation of the Regge slope since for the vector correlator there is a one by one correspondence between the dilaton modification coefficients and the OPE coefficients eq. (\ref{eq:Fixingbs}). We have illustrated here that the partial information of few OPE terms can be treated in our model such that the QCD properties on the Minkowski region are still satisfied, in other words, we assume Regge trajectories even if it is realized dynamically and impose a correct OPE.

\section{Diagrammatic representations of the results}
\label{app:diagrams}

In order to simplify the understanding of the calculation procedure, we can represent graphically the result by constructing diagrammatic rules. Let us  define  the  \textit{bulk-connectors} and  \textit{boundaries-connectors}, as illustrated on tables \ref{tab:FeynmanRules1} and \ref{tab:FeynmanRules2}. One can associate to each connector an integral and just by multiplying each of them, we obtain the corresponding contribution. 
The last rule is that to obtain all the contributions to $\Pi_V^{(n)}$, one needs to collect all the possible ways to have a path with $n$ legs starting from the boundary-connector. 

One can see all the possible combinations on figure \ref{fig:GraphicalPiV} for the vectorial case and on figure \ref{fig:GraphicalPiA} for the axial case. For example, it is easy to recover eq. (\ref{eq:Pi(1)fG}) using this rules.

\begin{table}[h]
\begin{center}
\begin{tabular}{p{6cm}l}
\begin{minipage}{6cm}
\tikzstyle{site}=[circle,draw=red!50,fill=red!20,thick]
\tikzstyle{place}=[circle,draw=blue!50,fill=blue!20,thick]
\tikzstyle{transition}=[rectangle,draw=black!50,fill=black!20,thick]
\begin{tikzpicture}[inner sep=2mm]	
  \node at (0,0) [place] (A){$b_{2k}$};
  \node at (2.1,0) (B) {};
  \node at (3.2,0) (C) {$k$ times};
  \draw [-,thick,blue!50] (A) -- (B);
  \draw [-,thick,blue!50] ([yshift=-0.6mm]B.west) -- ([yshift=-0.6mm]A.east);
  \draw [-,thick,blue!50] ([yshift=0.6mm]B.west) -- ([yshift=0.6mm]A.east);
  \draw [-,thick,blue!50] ([yshift=-1.2mm]B.west) -- ([yshift=-1.2mm]A.east);
  \draw [-,thick,blue!50,dashed] ([yshift=-2mm]B.west) -- ([yshift=-2mm]A.east);
  \draw [snake=brace, mirror snake,thick,blue!50] (2.4,-0.4) -- (2.4,0.4) ;
  \filldraw [black] (1.9,-0.05) circle (3.1pt);
  \filldraw [black] (-0.6,0) circle (1pt);
  \node at (-0.8,0)  {$x$};
  \node at (2.2,0)  {$z$};
\end{tikzpicture}
\end{minipage}
& 
 $\displaystyle = b_{2k} \int_0^ \infty \d x \; w_0(x) G_V(Q^2;z,x) x^k \partial_x f_V^{(0)}(Q^2,x)$\\
&\\
\begin{minipage}{6cm}
\tikzstyle{site}=[circle,draw=red!50,fill=red!20,thick]
\tikzstyle{place}=[circle,draw=blue!50,fill=blue!20,thick]
\tikzstyle{transition}=[rectangle,draw=black!50,fill=black!20,thick]
\begin{tikzpicture}[inner sep=2mm]	
  \node at (0,0) [place] (A){$b_{2k}$};
  \node at (2.1,0) (B) {};
  \node at (3.2,0) (C) {$k$ times};
  \node at (-2,0) (D) {};
  \draw [-,thick,blue!50] (A) -- (B);
  \draw [-,thick,blue!50] ([yshift=-0.6mm]B.west) -- ([yshift=-0.6mm]A.east);
  \draw [-,thick,blue!50] ([yshift=0.6mm]B.west) -- ([yshift=0.6mm]A.east);
  \draw [-,thick,blue!50] ([yshift=-1.2mm]B.west) -- ([yshift=-1.2mm]A.east);
  \draw [-,thick,blue!50,dashed] ([yshift=-2mm]B.west) -- ([yshift=-2mm]A.east);
  \draw [snake=brace, mirror snake,thick,blue!50] (2.4,-0.4) -- (2.4,0.4) ;
  \draw [-,thick,blue!50] (A) -- (D);
  \filldraw [black] (1.9,-0.05) circle (3.1pt);
  \filldraw [black] (-0.6,0) circle (1pt);
  \node at (-1,0.3) {$F(x)$};
  \node at (2.2,0)  {$z$};
\end{tikzpicture}
\end{minipage}
& 
 $\displaystyle = b_{2k} \int_0^ \infty \d x \; w_0(x) G_V(Q^2;z,x) x^k \partial_x F(x)$\\
&\\
\begin{minipage}{6cm}
\tikzstyle{site}=[circle,draw=red!50,fill=red!20,thick]
\tikzstyle{place}=[circle,draw=blue!50,fill=blue!20,thick]
\tikzstyle{transition}=[rectangle,draw=black!50,fill=black!20,thick]
\begin{tikzpicture}[inner sep=2mm]	
  \node at (0,0) [site] (A){$\beta_{2k}$};
  \node at (2.1,0) (B) {};
  \node at (3.2,0) (C) {$k$ times};
  \draw [-,thick,red!50] (A) -- (B);
  \draw [-,thick,red!50] ([yshift=-0.6mm]B.west) -- ([yshift=-0.6mm]A.east);
  \draw [-,thick,red!50] ([yshift=0.6mm]B.west) -- ([yshift=0.6mm]A.east);
  \draw [-,thick,red!50] ([yshift=-1.2mm]B.west) -- ([yshift=-1.2mm]A.east);
  \draw [-,thick,red!50,dashed] ([yshift=-2mm]B.west) -- ([yshift=-2mm]A.east);
  \draw [snake=brace, mirror snake,thick,red!50] (2.4,-0.4) -- (2.4,0.4) ;
  \filldraw [black] (1.9,-0.05) circle (3.1pt);
  \filldraw [black] (-0.6,0) circle (1pt);
  \node at (-0.8,0)  {$x$};
  \node at (2.2,0)  {$z$};
\end{tikzpicture}
\end{minipage}
& 
 $\displaystyle = g_5^2\beta_{2k} \int_0^ \infty \d x \; w_0(x) G_A(Q^2;z,x) x^k  f_A^{(0)}(Q^2,x)$\\
&\\
\begin{minipage}{6cm}
\tikzstyle{site}=[circle,draw=red!50,fill=red!20,thick]
\tikzstyle{place}=[circle,draw=blue!50,fill=red!20,thick]
\tikzstyle{transition}=[rectangle,draw=black!50,fill=black!20,thick]
\begin{tikzpicture}[inner sep=2mm]	
  \node at (0,0) [site] (A){$\beta_{2k}$};
  \node at (2.1,0) (B) {};
  \node at (3.2,0) (C) {$k$ times};
  \node at (-2,0) (D) {};
  \draw [-,thick,red!50] (A) -- (B);
  \draw [-,thick,red!50] ([yshift=-0.6mm]B.west) -- ([yshift=-0.6mm]A.east);
  \draw [-,thick,red!50] ([yshift=0.6mm]B.west) -- ([yshift=0.6mm]A.east);
  \draw [-,thick,red!50] ([yshift=-1.2mm]B.west) -- ([yshift=-1.2mm]A.east);
  \draw [-,thick,red!50,dashed] ([yshift=-2mm]B.west) -- ([yshift=-2mm]A.east);
  \draw [snake=brace, mirror snake,thick,red!50] (2.4,-0.4) -- (2.4,0.4) ;
  \draw [-,thick,red!50] (A) -- (D);
  \filldraw [black] (1.9,-0.05) circle (3.1pt);
  \filldraw [black] (-0.6,0) circle (1pt);
  \node at (-1,0.3) {$F(x)$};
  \node at (2.2,0)  {$z$};
\end{tikzpicture}
\end{minipage}
& 
 $\displaystyle = g_5^2\beta_{2k} \int_0^ \infty \d x \; w_0(x) G_A(Q^2;z,x) x^k F(x)$\\
&\\
\end{tabular}
\end{center}
\caption{Diagrammatic rules of construction: the bulk-connectors. On the right one has the corresponding operator.}\label{tab:FeynmanRules1}
\end{table}

\begin{table}[h]
\begin{center}
\begin{tabular}{p{6cm}l}
\begin{minipage}{6cm}
\tikzstyle{site}=[circle,draw=red!50,fill=red!20,thick]
\tikzstyle{place}=[circle,draw=blue!50,fill=red!20,thick]
\tikzstyle{transition}=[rectangle,draw=black!50,fill=black!20,thick]
\begin{tikzpicture}[inner sep=2mm]	
  \node at (0,0) [transition] (A){$\Pi_V^{(n)}$};
  \node at (1.8,0) (B) {};
  \draw [-,thick,blue!50] (A) -- (B);
  \filldraw [black] (0.55,0) circle (1pt);
  \node at (1.2,0.4)  {$F(z)$};
\end{tikzpicture}
\end{minipage}
& 
\begin{minipage}{7cm}
\begin{alignat*}{2}
= -\frac{1}{g_5^2} \int_0^ \infty \d z \; w_0(z) f^{(0)}_V(Q^2,z) \delta(z) \partial_z F(z)\\
= -\frac{1}{g_5^2} \lim_{z \rightarrow 0}  w_0(z) f^{(0)}_V(Q^2,z) \partial_z F(z)\hspace*{0.8cm}
\end{alignat*}
\end{minipage}\\
&\\
\begin{minipage}{6cm}
\tikzstyle{site}=[circle,draw=red!50,fill=red!20,thick]
\tikzstyle{place}=[circle,draw=blue!50,fill=red!20,thick]
\tikzstyle{transition}=[rectangle,draw=black!50,fill=black!20,thick]
\begin{tikzpicture}[inner sep=2mm]	
  \node at (0,0) [transition] (A){$\Pi_A^{(n)}$};
  \node at (1.8,0) (B) {};
  \draw [-,thick,blue!50] (A) -- (B);
  \filldraw [black] (0.55,0) circle (1pt);
  \node at (1.2,0.4)  {$F(z)$};
\end{tikzpicture}
\end{minipage}
& 
 \begin{minipage}{7cm}
\begin{alignat*}{2}
= -\frac{1}{g_5^2} \int_0^ \infty \d z \; w_0(z) f^{(0)}_A(Q^2,z) \delta(z) \partial_z F(z)\\
= -\frac{1}{g_5^2} \lim_{z \rightarrow 0}  w_0(z) f^{(0)}_A(Q^2,z) \partial_z F(z)\hspace*{0.8cm}
\end{alignat*}
\end{minipage}\\
&\\
\end{tabular}
\end{center}
\caption{Diagrammatic rules of construction: the boundary-connectors. On the right one has the corresponding operator.}\label{tab:FeynmanRules2}
\end{table}

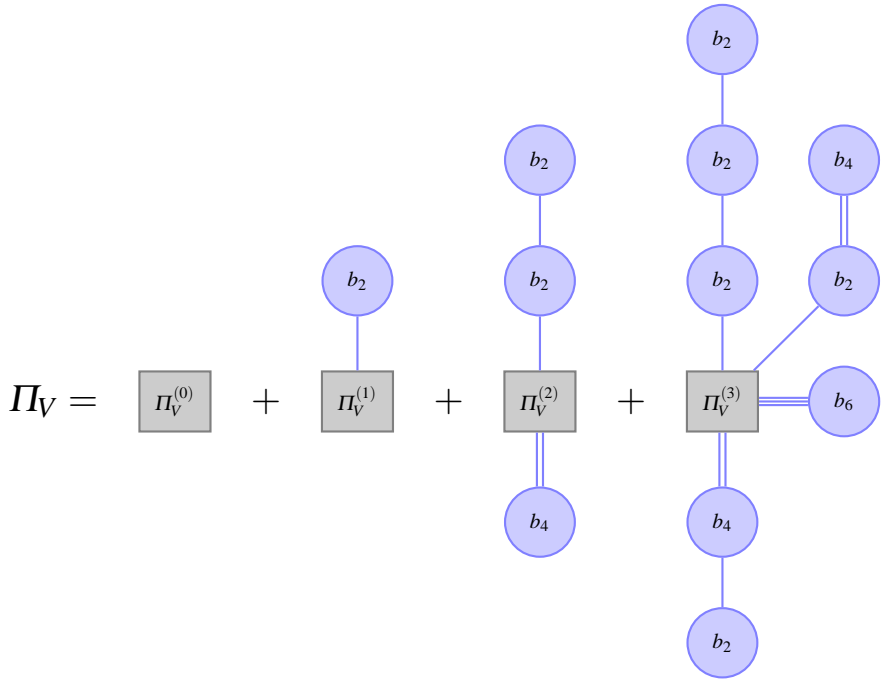
\begin{figure}[h]
\begin{center}
\tikzstyle{place}=[circle,draw=blue!50,fill=blue!20,thick]
\tikzstyle{transition}=[rectangle,draw=black!50,fill=black!20,thick]
\begin{tikzpicture}[scale=0.8,inner sep=2mm]
  \node at ( 0,2) [place] (A){$b_2$};
  \node at ( 3,2) [place] (B){$b_2$};
  \node at ( 3,4) [place] (C){$b_2$};
  \node at ( 3,-2) [place] (D){$b_4$};
  \node at ( 6,2) [place] (E){$b_2$};
  \node at ( 6,4) [place] (F){$b_2$};
  \node at ( 6,6) [place] (G){$b_2$}; 
  \node at ( 6,-2) [place] (H){$b_4$};
  \node at ( 6,-4) [place] (I){$b_2$}; 
  \node at ( 8,0) [place] (J){$b_6$};
  \node at ( 8,2) [place] (K){$b_2$};
  \node at ( 8,4) [place] (L){$b_4$};
  \node at (-5,0)   {\Large $\Pi_V=$};
  \node at (-3,0) [transition] (Pi0) {$\Pi_V^{(0)}$};
  \node at (-1.5,0)   {\Large $+$};
  \node at ( 0,0) [transition] (Pi1){$\Pi_V^{(1)}$};
  \node at (1.5,0)   {\Large $+$};
  \node at ( 3,0) [transition] (Pi2){$\Pi_V^{(2)}$};
  \node at (4.5,0)   {\Large $+$};
  \node at ( 6,0) [transition] (Pi3){$\Pi_V^{(3)}$};
  \draw [-,thick,blue!50] (A) -- (Pi1);
  \draw [-,thick,blue!50] (B) -- (Pi2);
  \draw [-,thick,blue!50] (C) -- (B);
  \draw [-,thick,blue!50] ([xshift=-0.5mm]D.north) -- ([xshift=-0.5mm]Pi2.south);
  \draw [-,thick,blue!50] ([xshift=0.5mm]D.north) -- ([xshift=0.5mm]Pi2.south);
  \draw [-,thick,blue!50] (E) -- (Pi3);
  \draw [-,thick,blue!50] (F) -- (E);
  \draw [-,thick,blue!50] (G) -- (F);
  \draw [-,thick,blue!50] ([xshift=-0.5mm]H.north) -- ([xshift=-0.5mm]Pi3.south);
  \draw [-,thick,blue!50] ([xshift=0.5mm]H.north) -- ([xshift=0.5mm]Pi3.south);  
  \draw [-,thick,blue!50] (I) -- (H);
  \draw [-,thick,blue!50] (Pi3) -- (J);
  \draw [-,thick,blue!50] ([yshift=-0.6mm]J.west) -- ([yshift=-0.6mm]Pi3.east);
  \draw [-,thick,blue!50] ([yshift=0.6mm]J.west) -- ([yshift=0.6mm]Pi3.east);
  \draw [-,thick,blue!50] (Pi3) -- (K);
  \draw [-,thick,blue!50] ([xshift=-0.5mm]K.north) -- ([xshift=-0.5mm]L.south);
  \draw [-,thick,blue!50] ([xshift=0.5mm]K.north) -- ([xshift=0.5mm]L.south);
\end{tikzpicture}
\caption{All graphical contributions to the vectorial two-point function up to $\Pi^{(3)}$}\label{fig:GraphicalPiV}
\end{center}
\end{figure}

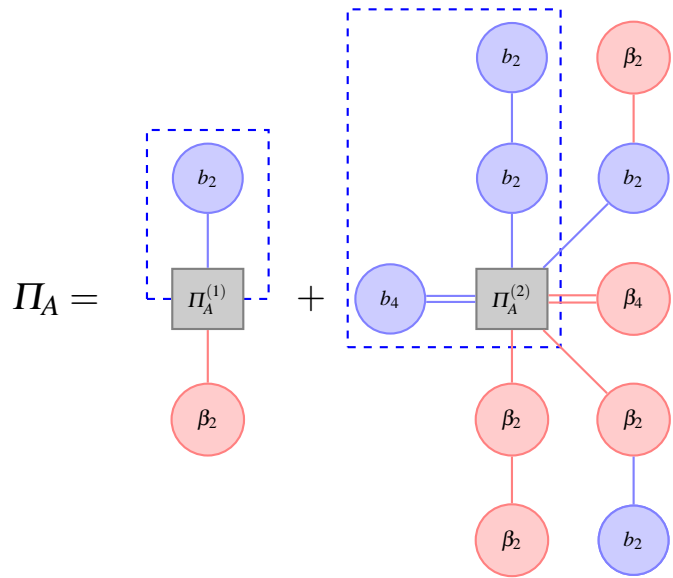
\begin{figure}[h]
\begin{center}
\tikzstyle{place}=[circle,draw=blue!50,fill=blue!20,thick]
\tikzstyle{site}=[circle,draw=red!50,fill=red!20,thick]
\tikzstyle{transition}=[rectangle,draw=black!50,fill=black!20,thick]
\begin{tikzpicture}[scale=0.8,inner sep=2mm]
  \node at ( -2,2) [place] (A){$b_2$};
  \node at ( 3,2) [place] (B){$b_2$};
  \node at ( 3,4) [place] (C){$b_2$};  
  \node at ( 1,0) [place] (D){$b_4$};
  \node at ( -2,-2) [site] (E){$\beta_2$};
  \node at ( 5,0) [site] (F){$\beta_4$};
  \node at ( 5,2) [place] (G){$b_2$};
  \node at ( 5,4) [site] (H){$\beta_2$};
  \node at ( 5,-2) [site] (I){$\beta_2$};
  \node at ( 5,-4) [place] (J){$b_2$};
  \node at ( 5,-4) [place] (K){$b_2$};
  \node at ( 3,-2) [site] (L){$\beta_2$};
  \node at ( 3,-4) [site] (M){$\beta_2$};
  \draw [thick,blue, dashed] (-3,0) rectangle (-1,2.8);
  \draw [thick,blue, dashed] (0.3,-0.8) rectangle (3.8,4.8);
  \node at (-4.5,0)   {\Large $\Pi_A =$};
  \node at ( -2,0) [transition] (Pi1){$\Pi_A^{(1)}$};
  \node at (-0.3,0)   {\Large $+$};
  \node at ( 3,0) [transition] (Pi2){$\Pi_A^{(2)}$};
  \draw [-,thick,blue!50] (A) -- (Pi1);
  \draw [-,thick,red!50]  (E) -- (Pi1);
  \draw [-,thick,blue!50] (B) -- (Pi2);
  \draw [-,thick,blue!50] (C) -- (B);
  \draw [-,thick,blue!50] ([yshift=-0.5mm]D.east) -- ([yshift=-0.5mm]Pi2.west);
  \draw [-,thick,blue!50] ([yshift=0.5mm]D.east) -- ([yshift=0.5mm]Pi2.west);
  \draw [-,thick,red!50] ([yshift=-0.6mm]F.west) -- ([yshift=-0.6mm]Pi2.east);
  \draw [-,thick,red!50] ([yshift=0.6mm]F.west) -- ([yshift=0.6mm]Pi2.east);
  \draw [-,thick,blue!50] (G) -- (Pi2);
  \draw [-,thick,red!50]  (H) -- (G);
  \draw [-,thick,red!50] (I) -- (Pi2);
  \draw [-,thick,blue!50] (J) -- (I);
  \draw [-,thick,red!50]  (L) -- (Pi2);
  \draw [-,thick,red!50]  (L) -- (M);
\end{tikzpicture}
\end{center}
\caption{The different contributions to the axial correlator up to $\Pi_A^{(2)}$. The part containing only the contributions from the shift are inside the dotted blue rectangle (\textit{i.e.} the path containing only blue lines).}\label{fig:GraphicalPiA}
\end{figure}

\clearpage



\begin{thebibliography}{99}

\bibitem{Peris:1998nj} 
S.~Peris, M.~Perrottet and E.~de Rafael,
"Matching long and short distances in large N(c) QCD",
JHEP {\bf 9805}, 011 (1998)

\bibitem{collins}
         P.D.B. Collins, 
         "Regge theory and particle physics",
	Phys. Reports {\bf 1} (1971) 103.


\bibitem{tHooft2d} 
G. 't Hooft, Nucl. Phys. B75 (1974)461;\\
C.G. Callan et al., Phys. Rev. D13 (1976) 1649;\\
M.B. Einhorn, Phys. Rev. D14 (1976) 3451.


\bibitem{Golterman:2001nk} 
  M.~Golterman and S.~Peris,
 "Large N(c) QCD meets Regge theory: The Example of spin one two point functions"
 JHEP {\bf 0101}, 028. 

  
\bibitem{Kaidalov:2001db} 
A.~B.~Kaidalov,
"Regge poles in QCD",
In Shifman, M. (ed.): At the frontier of particle physics, vol. 1 603-636

\bibitem{Shifman:2007xn} 
M.~Shifman and A.~Vainshtein,
"Highly Excited Mesons, Linear Regge Trajectories and the Pattern of the Chiral Symmetry Realization",
Phys.\ Rev.\ D {\bf 77}, 034002 (2008)


\bibitem{Maldacena:1997re}
J.~M.~Maldacena,
"The large N limit of superconformal field theories and supergravity",
Adv.\ Theor.\ Math.\ Phys.\  {\bf 2}, 231 (1998).

\bibitem{Gubser:1998bc} 
  S.~S.~Gubser, I.~R.~Klebanov and A.~M.~Polyakov,
  "Gauge theory correlators from noncritical string theory",
  Phys.\ Lett.\ B {\bf 428}, 105 (1998)
  
\bibitem{Witten:1998qj} 
  E.~Witten,
  "Anti-de Sitter space and holography",
  Adv.\ Theor.\ Math.\ Phys.\  {\bf 2}, 253 (1998)


\bibitem{Erlich:2005qh} 
  J.~Erlich, E.~Katz, D.~T.~Son and M.~A.~Stephanov,
  "QCD and a holographic model of hadrons",
  Phys.\ Rev.\ Lett.\  {\bf 95}, 261602 (2005)


\bibitem{Da Rold:2005zs} 
  L.~Da Rold and A.~Pomarol,
  "Chiral symmetry breaking from five dimensional spaces",
  Nucl.\ Phys.\ B {\bf 721}, 79 (2005)


\bibitem{Hirn:2005nr} 
  J.~Hirn and V.~Sanz,
  "Interpolating between low and high energy QCD via a 5-D Yang-Mills model",
  JHEP {\bf 0512}, 030 (2005)


\bibitem{Son:2003et} 
  D.~T.~Son and M.~A.~Stephanov,
  "QCD and dimensional deconstruction",
  Phys.\ Rev.\ D {\bf 69}, 065020 (2004)


\bibitem{Sakai:2004cn} 
  T.~Sakai and S.~Sugimoto,
  "Low energy hadron physics in holographic QCD",
  Prog.\ Theor.\ Phys.\  {\bf 113}, 843 (2005)

\cite{Brodsky:2014yha}
\bibitem{Brodsky:2014yha} 
  S.~J.~Brodsky, G.~F.~de Teramond, H.~G.~Dosch and J.~Erlich,
  "Light-Front Holographic QCD and Emerging Confinement",
  arXiv:1407.8131 [hep-ph].


\bibitem{Migdal:1977ut} 
  A.~A.~Migdal,
  "Series Expansion for Mesonic Masses in Multicolor QCD",
  Annals Phys.\  {\bf 110}, 46 (1978).


\bibitem{Shifman:2005zn} 
  M.~Shifman,
 "Highly excited hadrons in QCD and beyond", 
  hep-ph/0507246.


\bibitem{Karch:2006pv} 
  A.~Karch, E.~Katz, D.~T.~Son and M.~A.~Stephanov,
  "Linear confinement and AdS/QCD",
  Phys.\ Rev.\ D {\bf 74}, 015005 (2006).

\bibitem{Andreev:2006vy} 
  O.~Andreev,
  "1/q**2 corrections and gauge/string duality", 
  Phys.\ Rev.\ D {\bf 73}, 107901 (2006)


\bibitem{Cata:2006ak} 
  O.~Cata,
  "Towards understanding Regge trajectories in holographic QCD",
  Phys.\ Rev.\ D {\bf 75}, 106004 (2007)

\bibitem{Zuo:2008re} 
  F.~Zuo and T.~Huang,
 "Comments on the two-dimensional power correction in the soft wall model",
  Chin.\ Phys.\ Lett.\  {\bf 25}, 3601 (2008)

\bibitem{Csaki:2006ji} 
  C.~Csaki and M.~Reece,
  "Toward a systematic holographic QCD: A Braneless approach",
  JHEP {\bf 0705}, 062 (2007)

\bibitem{Veneziano:1968yb} 
  G.~Veneziano,
  "Construction of a crossing - symmetric, Regge behaved amplitude for linearly rising trajectories",
  Nuovo Cim.\ A {\bf 57}, 190 (1968).

\bibitem{Golterman:2001pj} 
  M.~Golterman, S.~Peris, B.~Phily and E.~De Rafael,
  "Testing an approximation to large N(c) QCD with a toy model",
  JHEP {\bf 0201}, 024 (2002)

\bibitem{Dominguez:2001zu} 
  C.~A.~Dominguez,
  "Pion form-factor in large N(c) QCD",
  Phys.\ Lett.\ B {\bf 512}, 331 (2001)

\bibitem{Afonin:2004yb} 
  S.~S.~Afonin, A.~A.~Andrianov, V.~A.~Andrianov and D.~Espriu,
  "Matching Regge theory to the OPE",
  JHEP {\bf 0404}, 039 (2004)

\bibitem{Cata:2005zj} 
  O.~Cata, M.~Golterman and S.~Peris,
  "Duality violations and spectral sum rules",
  JHEP {\bf 0508}, 076 (2005).

\bibitem{Kwee:2007nq} 
  H.~J.~Kwee and R.~F.~Lebed,
  "Pion Form Factor in Improved Holographic QCD Backgrounds",
  Phys.\ Rev.\ D {\bf 77}, 115007 (2008)

\bibitem{Gherghetta:2009ac} 
  T.~Gherghetta, J.~I.~Kapusta and T.~M.~Kelley,
  "Chiral symmetry breaking in the soft-wall AdS/QCD model",
  Phys.\ Rev.\ D {\bf 79}, 076003 (2009)

\bibitem{Beneke:1998ui} 
  M.~Beneke,
  "Renormalons",
  Phys.\ Rept.\  {\bf 317}, 1 (1999)


\bibitem{thooft}
         G. 't Hooft, 
         "A Planar Diagram Theory for Strong Interactions",
         Nucl. Phys. {\bf B72} (1974) 461.

\bibitem{witten}
         E. Witten,
         "Baryons in the 1/N Expansion",
          Nucl. Phys. {\bf B160} (1979) 57.


\bibitem{Beringer:1900zz} 
  J.~Beringer {\it et al.}  [Particle Data Group Collaboration],
  "Review of Particle Physics (RPP)",
  Phys.\ Rev.\ D {\bf 86}, 010001 (2012).


\bibitem{Masjuan:2012gc} 
  P.~Masjuan, E.~Ruiz Arriola and W.~Broniowski, "Systematics of radial and angular-momentum Regge trajectories of light non-strange $q\bar{q}$-states",
  Phys.\ Rev.\ D {\bf 85}, 094006 (2012)


\bibitem{SVZ}
       M. Shifman, A. Vainshtein and V. Zakharov,
       "QCD and Resonance Physics. The rho-omega Mixing",
       Nucl. Phys. {\bf B147} (1979) 385; 447.
       

\bibitem{Grigoryan:2007my} 
  H.~R.~Grigoryan and A.~V.~Radyushkin,
  "Structure of vector mesons in holographic model with linear confinement"
  Phys.\ Rev.\ D {\bf 76}, 095007 (2007).


\bibitem{Jamin:2011vd} 
  M.~Jamin,
  "What two models may teach us about duality violations in QCD",
  JHEP {\bf 1109}, 141 (2011).

\bibitem{deRafael:2012PJ} 
	E. de Rafael,
	"Large Nc QCD and Harmonic Sums"
	Pramana-journal of physics, Indian Academy of Sciences, Vol. 78, N6 June 2012 pp. 927-946.	

\bibitem{Mondejar:2008dt} 
  J.~Mondejar and A.~Pineda,
  "1/N(c) and 1/n preasymptotic corrections to Current-Current correlators",
  JHEP {\bf 0806}, 039 (2008)

\bibitem{CDAG02}
L. Cappiello, G. D'Ambrosio and D. Greynat,
"The pion wave function in a Soft Wall Model", 
\textit{To be published}

\bibitem{Cox:2014zea}
  P.~Cox and T.~Gherghetta,
  "A Soft-Wall Dilaton", 
  JHEP {\bf 1502} (2015) 006

\bibitem{Colangelo:2011xk} 
  P.~Colangelo, F.~De Fazio, J.~J.~Sanz-Cillero, F.~Giannuzzi and S.~Nicotri,
 "Anomalous $AV^*V$ vertex function in the soft-wall holographic model of QCD",
  Phys.\ Rev.\ D {\bf 85}, 035013 (2012)

\bibitem{Ecker:2013xja}
  G.~Ecker, 
  "Facets of chiral perturbation theory",
  Nucl.\ Phys.\ Proc.\ Suppl.\  {\bf 245} (2013) 1

\bibitem{deRafael:1995zv} 
  E.~de Rafael,
  "Chiral Lagrangians and kaon CP violation",
  In *Boulder 1994, Proceedings, CP violation and the limits of the standard model* 15-84, and Marseille Cent. Theor. Phys. - 95-P-3161 (95/01,rec.Feb.) 71 p. (507515)
  
\bibitem{Hirn:2005vk} 
  J.~Hirn, N.~Rius and V.~Sanz,
  "Geometric approach to condensates in holographic QCD",
  Phys.\ Rev.\ D {\bf 73}, 085005 (2006)

\bibitem{shi94}
M.~A. Shifman, {\it {Theory of pre-asymptotic effects in weak inclusive
  decays}}. In Proc. of the Workshop {\em Continuous advances in QCD},
  ed. A.~Smilga (World Scientific, Singapore, 1994).

\bibitem{shi95}
M.~A. Shifman, {\it {Recent progress in the heavy quark theory}},
 In {\em Particles, Strings and Cosmology}, eds. J.~Bagger et al. (World Scientific,
  Singapore, 1996).

\bibitem{cdsu96}
B.~Chibisov, R.~D. Dikeman, M.~A. Shifman, and N.~Uraltsev, {\it {Operator
  product expansion, heavy quarks, QCD duality and its violations}},  {\em Int.
  J. Mod. Phys.} {\bf A12} (1997) 2075--2133,.

\bibitem{bsz98}
B.~Blok, M.~A. Shifman, and D.-X. Zhang, {\it {An Illustrative example of how
  quark hadron duality might work}},  {\em Phys. Rev.} {\bf D57} (1998)
  2691--2700.

\bibitem{shi00}
M.~A. Shifman, {\it {Quark-hadron duality}}.
  Published in the Boris Ioffe Festschrift 'At the Frontier of Particle Physics Handbook of QCD', ed. M. Shifman (World Scientific, Singapore, 2001).

\bibitem{GonzalezAlonso:2010xf} 
  M.~Gonzalez-Alonso, A.~Pich and J.~Prades,
  "Pinched weights and Duality Violation in QCD Sum Rules: a critical analysis",
  Phys.\ Rev.\ D {\bf 82}, 014019 (2010)

\bibitem{Batell:2008zm} 
  B.~Batell and T.~Gherghetta,
  "Dynamical Soft-Wall AdS/QCD",
  Phys.\ Rev.\ D {\bf 78}, 026002 (2008)


\bibitem{deTeramond:2009xk} 
  G.~F.~de Teramond and S.~J.~Brodsky,
  "Light-Front Holography and Gauge/Gravity Duality: The Light Meson and Baryon Spectra",
  Nucl.\ Phys.\ Proc.\ Suppl.\  {\bf 199}, 89 (2010)


\bibitem{Cappiello:2009cj} 
  L.~Cappiello and G.~D'Ambrosio,
  "On the Evaluation of Gluon Condensate Effects in the Holographic Approach to QCD",
  Eur.\ Phys.\ J.\ C {\bf 69}, 315 (2010)

\bibitem{Domenech:2010aq} 
  O.~Domenech, G.~Panico and A.~Wulzer,
  "Massive Pions, Anomalies and Baryons in Holographic QCD",
  Nucl.\ Phys.\ A {\bf 853}, 97 (2011)
  

\bibitem{Gambino:2011cq} 
  P.~Gambino,
  "B semileptonic moments at NNLO", JHEP {\bf 1109}, 055 (2011)



\bibitem{Rainv:1960}
	E. D. Rainville, "Special Functions", Macmillan, NewYork, 1960.


\bibitem{Gotts:1988} 
	J. E. Gottschalk and E. N. Maslen,
	"Reduction formulae for generalised hypergeometric functions of one variable",
	J. Phys. A. Math. Gen. 21 (1988) 1983-1998.



\end{thebibliography}
\end{document}